\documentclass[fleqn,usenatbib,useAMS]{mnras}

\usepackage{graphicx}	
\usepackage{mwe}
\usepackage{amsmath}	
\usepackage{amssymb}	
\usepackage{multicol}        
\usepackage{bm}		

\usepackage{pdflscape}	
\usepackage{multirow}
\usepackage{ulem}
\usepackage{comment}
\usepackage{array}
\usepackage{caption}
\usepackage{subcaption}
\usepackage[switch, modulo]{lineno}
\usepackage{float}

\usepackage{scalerel}
\usepackage{tikz}
\usetikzlibrary{svg.path}
\includecomment{comment} 
\definecolor{orcidlogocol}{HTML}{A6CE39}
\tikzset{
  orcidlogo/.pic={
    \fill[orcidlogocol] svg{M256,128c0,70.7-57.3,128-128,128C57.3,256,0,198.7,0,128C0,57.3,57.3,0,128,0C198.7,0,256,57.3,256,128z};
    \fill[white] svg{M86.3,186.2H70.9V79.1h15.4v48.4V186.2z}
                 svg{M108.9,79.1h41.6c39.6,0,57,28.3,57,53.6c0,27.5-21.5,53.6-56.8,53.6h-41.8V79.1z M124.3,172.4h24.5c34.9,0,42.9-26.5,42.9-39.7c0-21.5-13.7-39.7-43.7-39.7h-23.7V172.4z}
                 svg{M88.7,56.8c0,5.5-4.5,10.1-10.1,10.1c-5.6,0-10.1-4.6-10.1-10.1c0-5.6,4.5-10.1,10.1-10.1C84.2,46.7,88.7,51.3,88.7,56.8z};
  }
}

\newcommand\orcidicon[1]{\href{https://orcid.org/#1}{\mbox{\scalerel*{
\begin{tikzpicture}[yscale=-1,transform shape]
\pic{orcidlogo};
\end{tikzpicture}
}{|}}}}

\newcommand{\abs}[1]{\lvert#1\rvert}

\usepackage[T1]{fontenc}
\usepackage{ae,aecompl}
\usepackage{newtxtext,newtxmath}
\usepackage{caption}
\usepackage{subcaption}
\usepackage{float}

\title[QPOs from Truncated, Tilted Discs]{Truncated, Tilted Discs as a Possible Source of Quasi-Periodic Oscillations}

\author[Bollimpalli et al]{D. A. Bollimpalli$^{\orcidicon{0000-0001-8835-8733}{1,2,3}}$\thanks{Contact e-mail: \href{mailto:deepika@mpa-garching.mpg.de}{deepika@mpa-garching.mpg.de}},
P. C. Fragile$^{\orcidicon{0000-0002-5786-186X}{3, 4}}$, J. W. Dewberry$^{\orcidicon{0000-0001-9420-5194}{5}}$, W. Klu\'zniak$^{\orcidicon{0000-0001-9043-8062}{6}}$
\\
$^{1}$Max-Planck-Institut f{\"u}r Astrophysik, 85741 Garching b. M{\"u}nchen, Germany\\
$^{2}$Center for Interdisciplinary Exploration \& Research in Astrophysics (CIERA), Physics \& Astronomy, Northwestern University, Evanston, IL 60202, USA\\
$^{3}$Department of Physics and Astronomy, College of Charleston, Charleston, SC 29424, USA\\
$^{4}$Kavli Institute for Theoretical Physics, Kohn Hall, University of California, Santa Barbara, CA 93107, USA \\
$^{5}$Canadian Institute for Theoretical Astrophysics, 60 St. George Street, Toronto, ON M5S 3H8, Canada\\
$^{6}$Nicolaus Copernicus Astronomical Center, ul. Bartycka 18, PL 00-716 Warsaw, Poland}
\date{}

\pubyear{2020}
\begin{document}
\label{firstpage}
\pagerange{\pageref{firstpage}--\pageref{lastpage}}
\maketitle

\begin{abstract}
Many accreting black holes and neutron stars exhibit rapid variability in their X-ray light curves, termed quasi-periodic oscillations (QPOs). The most commonly observed type is the low-frequency ($\lesssim 10$ Hz), type-C QPO, while only a handful of sources exhibit high frequency QPOs ($\gtrsim 60$ Hz). The leading model for the type-C QPO is Lense-Thirring precession of a hot, geometrically thick accretion flow that is misaligned with the black hole's spin axis. However, existing versions of this model have not taken into account the effects of a surrounding, geometrically thin disc on the precessing, inner, geometrically thick flow. In Bollimpalli et. al 2023, using a set of GRMHD simulations of tilted, truncated accretion discs, we confirmed that the outer thin disc slows down the precession rate of the precessing torus, which has direct observational implications for type-C QPOs. In this paper, we provide a detailed analysis of those simulations and compare them with an aligned truncated disc simulation. We find that the misalignment of the disc excites additional variability in the inner hot flow, which is absent in the comparable aligned-disc simulations. This suggests that the misalignment may be a crucial requirement for producing QPOs. We attribute this variability to global vertical oscillations of the inner torus at epicyclic frequencies corresponding to the transition radius. This explanation is consistent with current observations of higher frequency QPOs in black hole X-ray binary systems. 
\end{abstract}



\section{Introduction}
\label{sec:intro}
Quasi-periodic oscillations (QPOs) are among the most intriguing phenomena observed in the X-ray light curves of accreting black hole and neutron star X-ray binary systems. In black holes these features are identified by relatively broad peaks in the power spectrum, where the broadening may result from the modulation of the oscillation frequency itself, or the finite lifetime of the oscillation. The frequency of these QPOs is believed to be closely linked to the strong gravity of the compact object \citep{Klis00, Klu01,Klis04, IM2019}. Therefore, understanding the physical nature of QPOs can be a step towards a better understanding of compact objects. Furthermore, timing studies probe smaller physical scales (of order the size of the compact object) than is possible with direct imaging of black hole binaries---only in the case of nearby supermassive black holes can such scales be probed with the current generation of the Event Horizon Telescope (EHT). 

In addition to the rapid variability, a fair population of neutron star and black hole X-ray binaries also exhibit ``state transitions", a phenomenon in which the spectral state of the source evolves concurrently with its luminosity, often tracing a ``q''-shape in a hardness-intensity diagram  \citep{NMY1996, EMN1997, B2005, MR2006, BA2014}. The two major spectral states observed in these systems, along with the various intermediate states, are the ``soft'' and ``hard'' states. In the soft state, the X-ray spectra are predominantly composed of a soft, thermal component (below approximately 3 keV). This soft emission likely arises from a relatively cold, optically thick and geometrically thin disc, exhibiting thermal blackbody-like radiation. On the other hand, in the hard state, the spectra are dominated by a power-law component (extending roughly between 10 - 100 keV). This power-law emission results from the Compton upscattering of soft seed photons emitted from the disc by a radiatively inefficient cloud of hot electrons, often referred to as the ``corona.'' The location and geometry of the corona are still the subject of considerable debate; some models suggest that it resides at the base of the jet \citep{MM1996, Fabian2009, Parker2015}, while others consider it to sandwich the disc \citep{GRV1979, HFM1993, S1996, B1999}. Alternatively, the truncated-disc model envisions a third possible geometry with a standard-thin disc truncated well outside the last stable orbit, and the corona as a geometrically thick, radiatively inefficient accretion flow filling the inner gap \citep{EL1975, EMN1997, LTMM2007}. Despite the uncertainty regarding the location of the corona, all of these models can reproduce the observed X-ray spectra during the ``state transitions" by associating them with appropriate changes in the geometry and strength of the accretion disc and corona, as well as the presence or absence of other physical components such as a jet.

QPOs of various flavours are observed during the different spectral states of a given source. Based on the frequencies, they are broadly classified into high-frequency ($\gtrsim 60$ Hz) and low-frequency ($\lesssim 30$ Hz) QPOs \citep{Remillard06, B2010}. High-frequency QPOs (HFQPOs) are typically observed in the high-flux/luminous accretion states \citep{BSM12, Motta16}. If occurring in the thin disc, they may be associated with trapped gravity ($g$-) and corrugation ($c$-) modes within the inner regions of the disc \citep{KF80, W1999,SWM01}, or the accretion disc responding quasi-periodically to a wobbling jet through the propagation of fast-magnetosonic waves~\citep{FM2022}, or the oscillations of spiral density waves moving between the disc's inner radius and the Lindblad radius at some frequency \citep{TP99, RVT2002}. If they occur in the geometrically thick (toroidal) flow, the HFQPOs may correspond to modes occurring at combinations of the orbital and epicyclic frequencies at characteristic radii  \citep{RYZ2003, BAKK2004, Blaes06,FSB16} or other gravity/inertial pressure modes within the accretion disc \citep{ABKR06}.

Low-frequency QPOs (LFQPOs) are further classified into type-A, B \& C, of which the first two are observed in the intermediate states during the transition to the soft spectral state \citep{Motta16}. Type-C QPOs are commonly observed in most accretion states~\citep{Motta12, Motta16}, although they are particularly prominent in the hard spectral state. For these QPOs, both the fractional QPO amplitude and the phase-lag between the hard and soft photons, measured at the QPO frequency, exhibit strong dependence on the inclination angle \citep{Motta2015, eijnden2017}. This inclination dependence provides compelling evidence for a geometric origin for this QPO. For a more comprehensive review on type-C QPOs, we recommend \citet{IM2019}.
Most of the models invoked to explain LFQPOs are based either on the geometry or the intrinsic properties of the accretion disc. Several potential mechanisms have been suggested, including oscillations of a standing shock in the accretion flow have been proposed as a source of LFQPOs. However, for such a  shock to generate the required frequencies, it would need to be located at a considerable distance from the black hole \citep{CT95}. 

Another popular model is the relativistic precession model, which associates observed QPOs with different frequencies derived from geodesic particle motion in general relativity, notably, nodal and periastron precession, as well as the particle orbit itself \citep{SV98, SVM99}. A variant of this model, which appears to align well with the observations of the type-C QPO, entails the precession of a finite inner region of the accretion flow instead of an infinitesimal particle or ring \citep{Ingram09}. This model corresponds with the truncated disc framework of the hard accretion state, as the precession is attributed to the hot, thick flow within the truncation radius (i.e., the corona) \citep{ID2011}. This model offers several advantages, including lowering the precession frequency into the appropriate range, naturally explaining the rise in QPO frequency during the early phases of an outburst cycle (when the truncation radius is moving inwards), and readily reproducing the inclination dependence observed in QPO studies (since precession is more apparent the closer to the orbital plane one observes). Furthermore, this model inherently accounts for the observed correlation between the QPO phase and the centroid of the iron emission line, when studied with phase-resolved spectroscopy \citep{Ingram2016}.

Recently, the rigid-body precession model of the torus has been extended to also explain the HFQPOs with a 3:2 ratio that are occasionally observed in the same sources \citep{FSB16}. In this extended framework, the HFQPOs are manifestations of natural, global oscillation modes of the same hot, thick flow that is undergoing Lense-Thirring precession. For both the HFQPOs and type-C QPO, the location of the truncation radius plays a critical role in setting their frequencies. During an X-ray outburst, as the source luminosity increases, the truncation radius moves inward, causing all QPO frequencies to rise. Eventually, the QPOs vanish when the truncation radius reaches the last stable orbit in the soft state. However, it is essential to note that the key assumption of this QPO model, particularly in relation to the type-C QPO, is that the inner, hot, and geometrically thick accretion flow can precess as a rigid body independently of the surrounding thin disc.

Previous numerical work employing magnetohydrodynamic simulations of {\it isolated}, tilted thick discs have demonstrated global Lense-Thirring precession at frequencies consistent with observations \citep{Fragile07, Liska18, WQB19}. Additionally, independent analytical and numerical studies have revealed that isolated gas tori exhibit global oscillation modes that match the general properties of observed HFQPOs \citep{Blaes06, Mishra17}. However, none of these studies have taken into account the full truncated-disc geometry, especially the impact of the surrounding geometrically thin disc on the QPO frequencies. A particular concern is understanding how the outer, thin disc affects the precession of the inner torus. To address this, we conducted a comprehensive set of 3D general relativistic magnetohydrodynamics (GRMHD) simulations of tilted and truncated discs around black holes using \textit{Cosmos++} \citep{Anninos05}. Through these simulations, we demonstrated a significant decrease in the precession rate of the inner torus due to the exchange of angular momentum between the inner torus and the outer thin disc \citep{BFK22}. This finding bears substantial implications for studies utilizing QPO frequencies to infer the truncation radius, as well as the mass and spin of the black hole. The purpose of this paper is to delve into further details regarding these simulations, particularly focusing on the properties of the accretion flow and the presence of additional QPO-like features. In addition to the simulations of tilted, truncated flows from \citet{BFK22}, we include a simulation of an aligned, truncated flow around a black hole to facilitate direct comparison.

The remainder of this paper is organized as follows: In Section~\ref{sec:setup}, we describe the initial setup of our simulations, including both aligned and misaligned configurations with respect to the black hole's spin axis. We then delve into the results of our simulations, particularly focusing on the properties and dynamics of the accretion disc (Section~\ref{sec:results}). In Section~\ref{sec:variability} we present the temporal variability observed in our simulations. Subsequently, in Section~\ref{sec:discussion}, we provide a contextual framework for our findings by comparing them to previous works and establishing their relevance to observations. Finally, we provide our concluding remarks in Section~\ref{sec:conclusions}.

\begin{figure*}
\begin{subfigure}[h!]{0.6\textwidth}
         \centering
         \includegraphics[width=\textwidth]{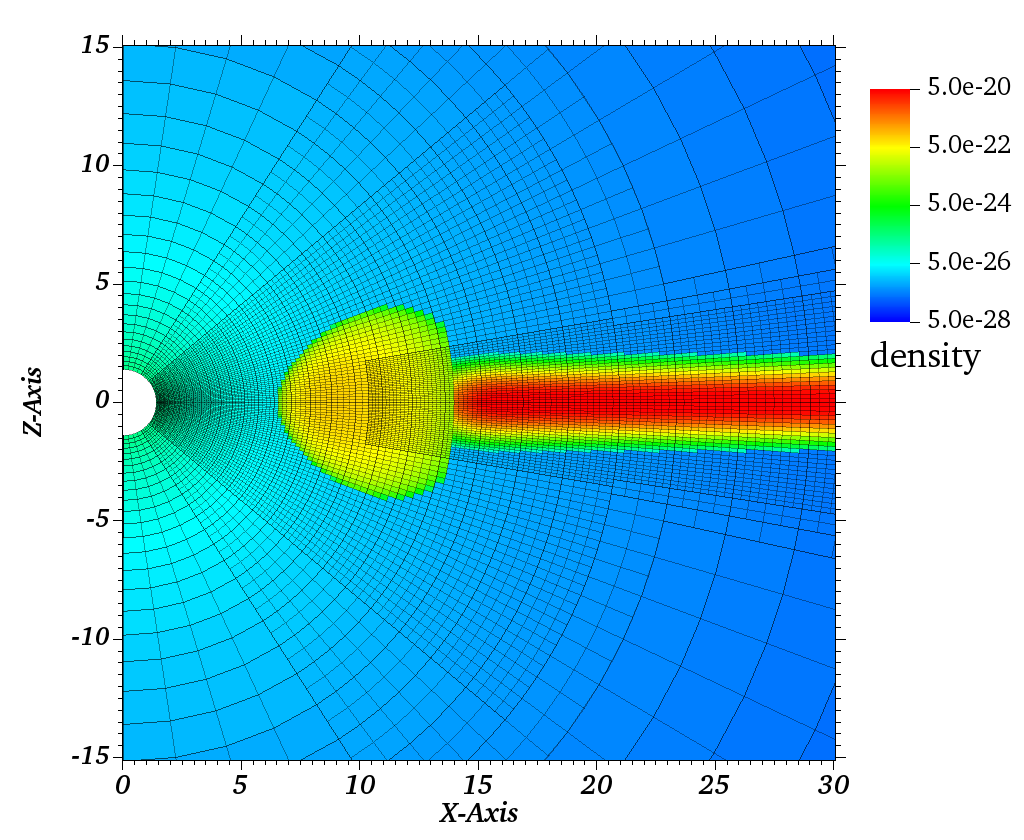}
\end{subfigure}
\begin{subfigure}[h!]{\textwidth}
         \centering
         \includegraphics[width=\textwidth]{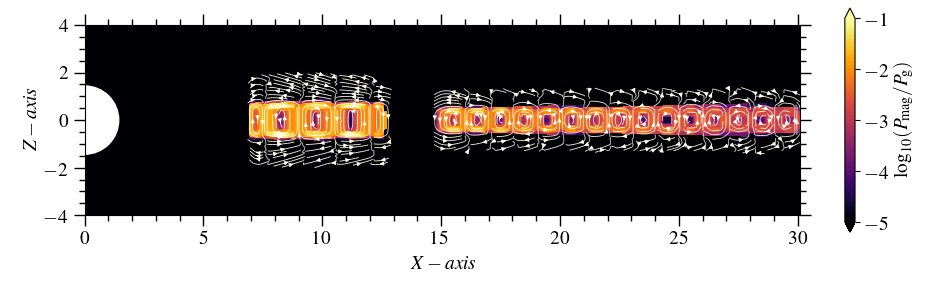}
\end{subfigure}
\caption{Initial setup. Top: Pseudocolour plot of gas density at $t=0$, overlaid with our 4-level mesh. Bottom: Pseudocolour plot of the ratio of magnetic pressure to gas pressure overlaid with magnetic field contours. The units of the axes are in $GM/c^2$, while the density is plotted in code units. }
\label{fig:setup}
\end{figure*}

\section{Initial Setup}
\label{sec:setup}
\subsection{Physical setup}
We conducted four, fully 3D GRMHD simulations of truncated accretion discs, as listed in Table 1: Two high-resolution, 4-level simulations, aimed at studying tilted and untilted scenarios, and two 3-level, low-resolution simulations to explore different black hole spins. The two 4-level simulations, a9b0L4 and a9b15L4, differed in terms of the initial alignment of angular momentum axes between the disc and the black hole (with dimensionless spin parameter $a_* = a/M = 0.9$ in both cases). In the former, the axes were aligned, while in the latter, they were misaligned by an angle $\beta_0 = 15^{\circ}$. In the tilted disc simulations, the initial disc angular momentum vector was oriented along the $z$-axis, while the black hole spin axis was tilted in the $-x$-direction. The 3-level simulations, a5b15L3 and a9b15L3, were both tilted $15^\circ$ and varied only in their black hole spins, with $a_* =0.5$ and 0.9, respectively.
\begin{center}
\begin{table*}
\begin{tabular*}{1 \textwidth}{cccccc}
\cline{1-6}
Simulation & Spin  &  Tilt & Resolution & $t_\mathrm{end}$  & $\Delta t_\mathrm{dump}$ \\
  & ($a_*$) & ($\beta_0$) & ($r\times \theta \times \phi)$ & [$GM/c^3$] & [$GM/c^3$]\\
\cline{1-6}
a9b0L4 & 0.9 & 0 & 384x256x256 & 62,250 & 300 {\rm and} 50 \\
a9b15L4 & 0.9 & $15^{\circ}$ & 384x256x256 & 25,000 & 10\\
a9b15L3 & 0.9 & $15^{\circ}$ & 192x128x128 & 15,000 &10\\
a5b15L3 & 0.5 & $15^{\circ}$ & 192x128x128  & 17,700 &10\\
\cline{1-6}
\end{tabular*}
\caption{Simulation parameters}
\label{tab:sims}
\end{table*}
\end{center}

The initial setup of all simulations comprises a finite torus surrounded by a thin disc, as depicted in Figure~\ref{fig:setup}. The torus is initialized using the method described by \citet{C1985}, assuming a constant specific angular momentum. An adiabatic equation of state is assumed with a polytropic index of $\gamma=5/3$. We chose the inner edge of the torus at $r_{\rm in} = 6.5\,GM/c^2$ and pressure maximum at $r_{\rm cen} = 9\,GM/c^2$. The torus is surrounded by a thin slab with a fixed height of $H = 0.4\,GM/c^2$, extending from $15\,GM/c^2$ to the outer boundary of the simulation domain at approximately $250\,GM/c^2$. The density distribution within the slab is determined by
\begin{equation}
\rho(R,z) \propto \frac{\exp[-z^2/(2H^2)]}{\{1+\exp[(15r_{\rm g}-R)/H]\}\{1+\exp[(R-(40 r_{\rm g})^{1.5})/H]\}} ~,
\end{equation}
where $R=r\sin \theta$ is the cylindrical radius and $ r_{\rm g} = GM/c^2$. The angular momentum within the disc follows a Keplerian distribution. To preserve the desired thin structure of the outer disc, we implement an artificial cooling function (only in the thin disc region), as outlined in \citet{FW12}, which is applied exclusively to radii beyond $15\,GM/c^2$. 

Both the torus and the slab are threaded with numerous small poloidal loops of magnetic field with alternating polarity, starting from a magnetic vector potential of the form:
\begin{equation}
    A_{\phi} \propto \frac{1}{1+\exp(\delta)}\sqrt{P_{\rm g}}\sin(0.4\mathrm{\pi} R/H) ~,
\end{equation} 
where $P_{\rm g}$ represents the local gas pressure and $\delta = 10 [(z/H)^2+ H^2/(R-r_{\rm ms})^2 + H^2/(40^{1.5}-R)^2-1]$. Here $H = 0.6$ and $0.4\,GM/c^2$ in the torus and disc regions, respectively, while $r_{\rm ms}$ signifies the location of the marginally stable orbit. The resulting magnetic field is normalized such that the initial ratio of the gas pressure to the magnetic pressure is $\ge 10$ throughout both the torus and disc, as shown in Fig.~\ref{fig:setup}. This magnetic field configuration impedes the accumulation of strong net flux in the inner regions, thus ruling out the formation of a magnetically arrested disc (MAD), which is appropriate for the luminous hard state \citep{Fragile23}. The background region surrounding the disc and torus is initialized with a non-magnetic gas with low-density, $\rho = 10^{-5}\rho_{\rm max}r^{-1.5}$, and internal energy density, $e = 10^{-7}e_{\rm max}r^{-2.5}$. Here $\rho_{\rm max}$ and $e_{\rm max}$ correspond to the maximum gas density and internal energy density within the disc/torus.

\subsection{Numerical setup}
All simulations are carried out using \textit{Cosmos++} \citep{Anninos05}, employing its five-stage, strong-stability-preserving Runge-Kutta (SSPRK) time-stepping scheme \citep{Spiteri02} and third-order, piecewise-parabolic spatial reconstruction \citep{Colella1984}. The fluxes at zone interfaces are computed using a two-wave, HLL (Harten-Lax-van Leer), approximate Riemann solver \citep{Harten1983}.

The simulations are performed in modified Kerr-Schild spherical-polar coordinates,  enabling the placement of the inner domain boundary inside the black hole event horizon. The grid is logarithmically spaced in $r$, covering the range $r\in [1.4 \,r_{\rm g},\,40^{1.5}\,r_{\rm g}]$, and uniformly spaced in $\phi$, encompassing the entire $[0,2\pi]$ domain. The polar angle $\theta$ is uniformly spaced in the coordinate $x_2$, with the grid spacing defined as
\begin{equation}
  \theta = \pi x_2 + \frac{1-h}{2} \sin(2 \pi x_2) ~,
\end{equation}
where $h=0.5$ is used to concentrate the grid cells close to the midplane \citep{McKinney06}. To avoid computing metric terms at the poles, a small cone of opening angle $10^{-15}\pi$ is excised.

All simulations have a base resolution of $48\times32\times32$, accompanied by two or three additional levels of static mesh refinement. Adjacent refinement levels differ in resolution by a factor of 2 in each dimension, resulting in a fiducial resolution of either $192\times128\times128$ or $384\times256\times256$ for most of the grid. The initial two levels of refinement are utilized to adequately resolve both the torus and the thin slab, which is common for both the 3-level and 4-level simulations. In the 4-level simulations, an additional refinement layer is implemented to enhance the resolution of the thin disc over the range $r\in[10.5,r_{\rm g}, 60,r_{\rm g}]$. In the untilted disc simulation, the region within $1.4<r/r_{\rm g}<21$ and $0.28\pi<\theta<0.72\pi$ undergoes two levels of refinement. However, for the tilted disc simulation, the second level of refinement for the torus region is adjusted to $1.4<r/r_{\rm g}<24$ and $0.2\pi<\theta<0.8\pi$. The thin disc region is refined by incorporating two levels of refinement within $12.6<r/r_{\rm g}<253$ and $0.45\pi<\theta<0.55\pi$, with the possibility of a third layer, as described earlier.

Outflow boundary conditions\footnote{All fields are copied to ghost zones, ensuring that the velocity component normal to the boundary points outward.} are applied at both the inner and outer radial boundaries. Near the poles, pole axis boundary conditions are utilized, where information from the corresponding zone across the pole is used for calculating gradients, although fluxes, magnetic fields, and electric fields (emf's) are set to zero on the faces and edges that touch the pole. Additionally, periodic boundary conditions are implemented along the azimuthal direction. To maintain the zero-divergence of the magnetic fields, the vector potential method \citep{FNSA18} is employed in the untilted disc simulation, while in the tilted disc simulations, the constrained transport scheme \citep{Fragile12} is utilized. Whenever density or pressure is boosted as a result of floors and ceilings, the boost is performed in the drift frame to preserve momentum along the magnetic field lines \citep{Ressler2017}.

As expected, once the simulations are initialized and evolve, the initial magnetic fields trigger the magnetorotational instability, leading to turbulence that facilitates the transport of angular momentum and enables accretion. The energy generated in this process causes the torus to inflate, resulting in the formation of a hot and thick flow in the inner regions. The cooling function employed in the thin slab region ($r \ge 15\,GM/c^2$) acts as a heat sink and helps to regulate the thickness of the disc at the target value of $H/r \approx 0.05$. The simulations have been run for over $25,000 \, GM/c^3$ for the 4-level simulations and approximately $15,000\,GM/c^3$ for the 3-level simulations. In all cases, the simulations successfully evolve into a two-component flow, as anticipated.
\section{Results}
\label{sec:results}
At a qualitative level, the results of both the 4-level and 3-level simulations exhibit similarities, which is promising as it indicates that our final conclusions are independent of the resolution. A significant finding from this study is that the simulations of tilted and untilted discs demonstrate notable and intriguing differences in both their dynamics and variability characteristics. These distinctions will be thoroughly explored in the subsequent two sections.

\subsection{Accretion flow properties}
\label{sec:accretion}
We initially aim to confirm that the simulations exhibit the desired two-component structure. One approach is to examine the density-squared weighted scale height of the disc, computed as:
\begin{equation}
H(r,t)/r = \left( \frac{\int \int  \rho^2 |\theta - \theta_{\rm mid}|^2 \sqrt{-g} \ d\theta \ d\phi}{\int \int \ \rho^2 \sqrt{-g} \ d\theta \ d\phi} \right)^{1/2} ~,
\label{eq:HoverR}
\end{equation}
where $\rho$ denotes the fluid density, $g$ is the metric determinant, and $\theta_{\text{mid}}$ represents the angular position of the disc's mid-plane. In the case of an untilted disc, $\theta_{\text{mid}}$ is taken to be $\pi/2$. However, in the case of a precessing disc, since the disc mid-plane departs from the symmetry plane of the grid, we need an alternate procedure. We choose to transform disc mid-plane back to the $x$-$y$ coordinate plane using the coordinate transformation described in \citet{WQB19}. In the remainder of this paper, whenever used, we denote the quantities in these transformed coordinates by a prime symbol ($'$).

\begin{figure}
\begin{subfigure}[h!]{0.46\textwidth}
         \centering
         \includegraphics[width=\textwidth]{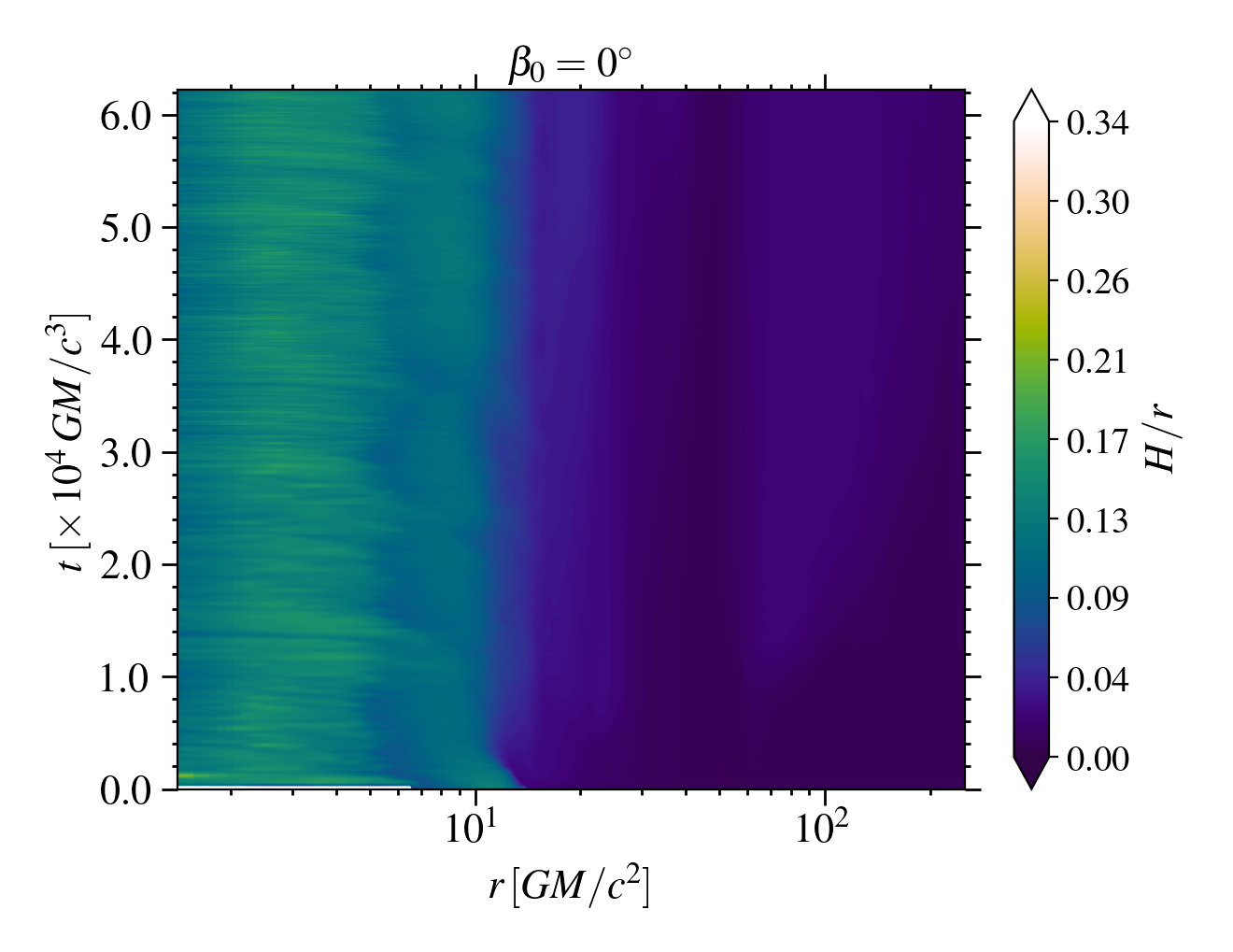}
\end{subfigure}
\begin{subfigure}[h!]{0.46\textwidth}
         \centering
         \includegraphics[width=\textwidth]{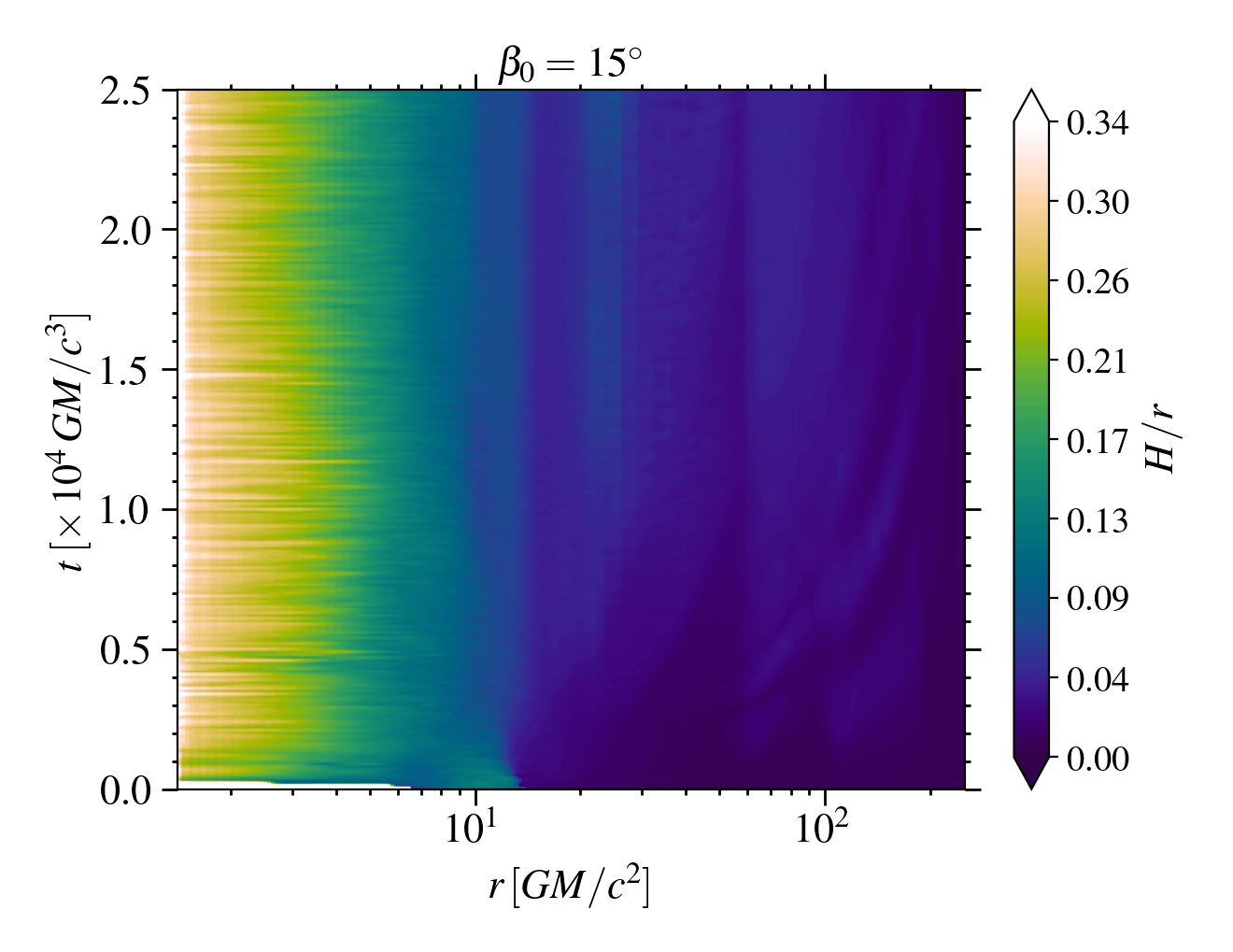}
\end{subfigure}
\caption{Space-time plot of the disc scale height, $H/r$, for the 4-level untilted (a9b0L4, top) and tilted (a9b15L4, bottom) disc simulations. The two-flow geometry is maintained well throughout the evolution in both cases.}
\label{fig:height}
\end{figure}

Fig.~\ref{fig:height} illustrates the evolution of the disc scale height as a function of radius and time for both the untilted and tilted disc simulations. In both cases, a distinct two-component flow is evident, comprising a geometrically thick disc in the inner region, surrounded by a geometrically thinner disc. The geometrically thin disc beyond $15\,GM/c^2$ maintains a scale height close to $0.05$ throughout the entire simulation, confirming that our cooling function in the thin disc region ($>\,15\,GM/c^2$) operates as expected. Also, the inner thick flow, which is at least three times thicker than the outer thin disc, maintains its scale height due to the advective nature of the thick accretion flows; most of the dissipated energy is advected inwards along with the matter (and eventually onto the black hole) before altering the vertical structure of the flow locally. Note that, in the tilted disc simulations, the scale height of the torus region is noticeably thicker compared to the untilted disc simulation. This increase in scale height is likely a consequence of additional dissipation associated with tilted discs \citep{Dexter13}, which agrees with our observation of at least an order of magnitude higher temperatures in the tilted simulation (a9b15L4) compared to the untilted one (a9b0L4). 

\begin{figure}
\begin{subfigure}[t]{0.48\textwidth}
         \centering
         \includegraphics[width=\textwidth]{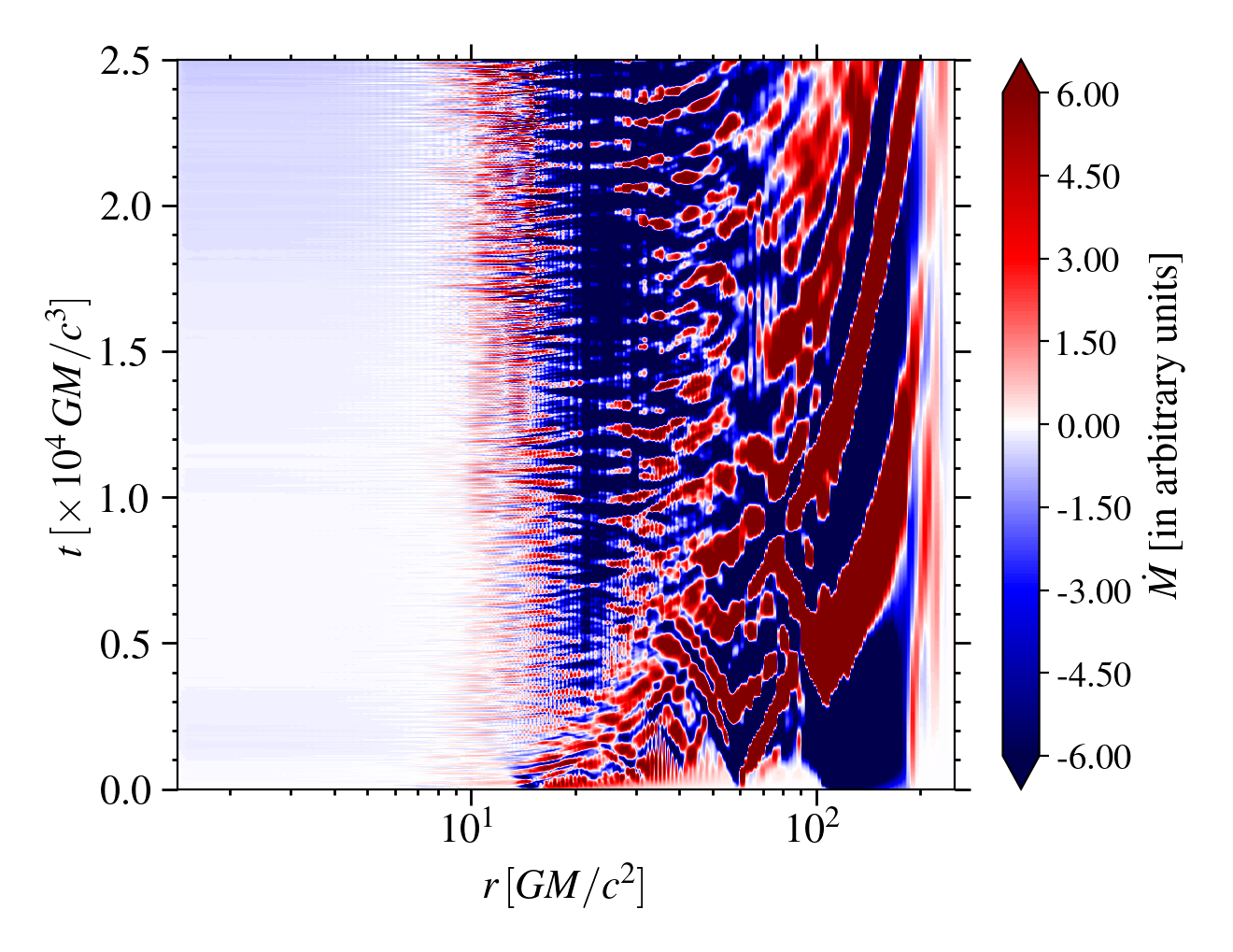}
\end{subfigure}
\begin{subfigure}[t]{0.48\textwidth}
         \centering
         \includegraphics[width=\textwidth]{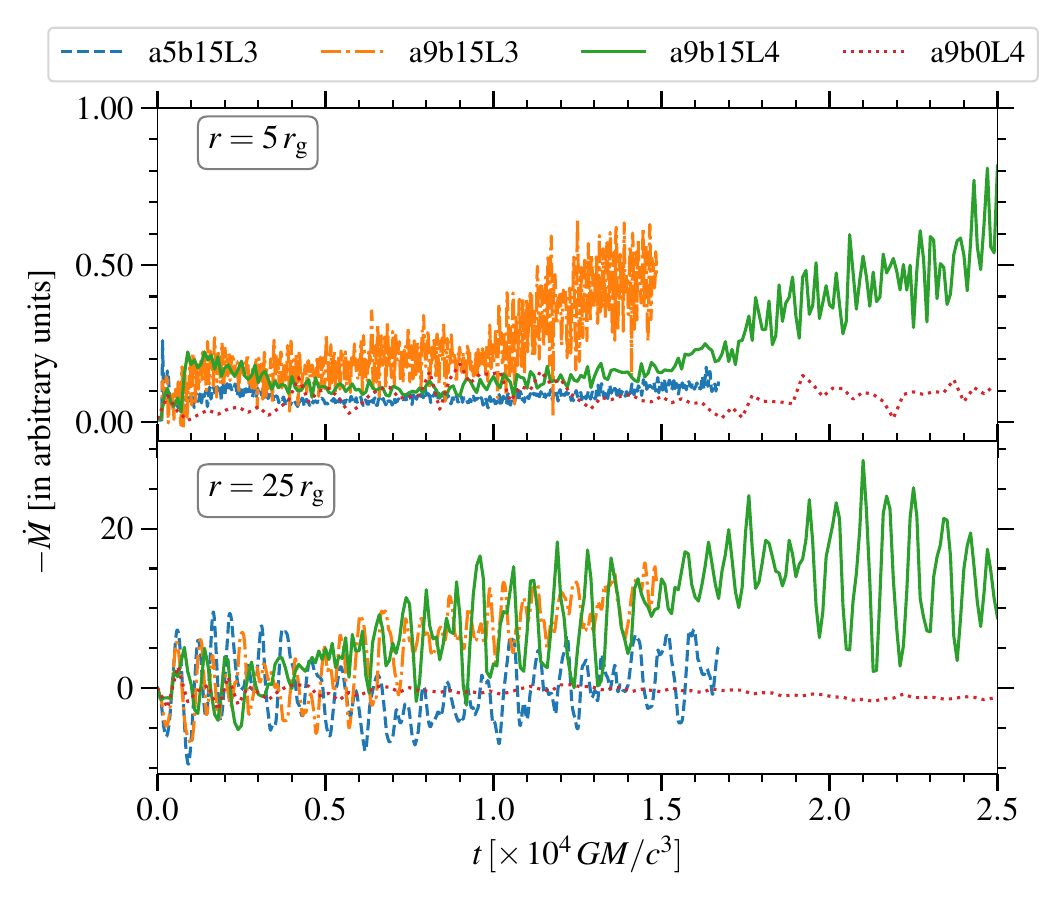}
\end{subfigure}
\caption{Top: Space-time plot of the mass accretion rate for simulation a9b15L4, measured in arbitrary units. Bottom: Evolution of the mass accretion rate at two radii, 5 and $25\,GM/c^2$, representing the thick- and thin-disc regions, for all four simulations. Note the differing scales of the two plots.}
\label{fig:mdot}
\end{figure}

In the top panel of Fig.~\ref{fig:mdot}, we show the time evolution of the mass accretion rate, measured in arbitrary units, for the high-resolution tilted simulation, a9b15L4. We define this quantity as
\begin{equation}
    \dot{M}(r,t) = \int \int \rho u^r \sqrt{-g} {\rm d}\theta {\rm d} \phi ~,
\end{equation}
such that a negative sign implies an inflow of matter toward the black hole. Here $u^r$ is the radial component of the fluid 4-velocity, $u^{\alpha}$. In general, the accretion rate in the thick disc region is at least an order of magnitude lower than that in the outer, geometrically thin region. The mismatch in $\dot{M}$ implies that material must be accumulating near the transition region (roughly between 11 and 19 $GM/c^2$), which is consistent with our findings of increased surface density at those radii (see for e.g., Fig~\ref{fig:sigma}). The mass accretion rate gradually increases in both the thick and thin-disc regions throughout the simulation. Similar results are observed in our other tilted disc simulations, as well. By closely comparing this figure to the bottom panel of Fig.~\ref{fig:height}, one can find correlation between the changes in the accretion rate and the scale height.

\begin{figure*}
\centering
\includegraphics[width= \textwidth]{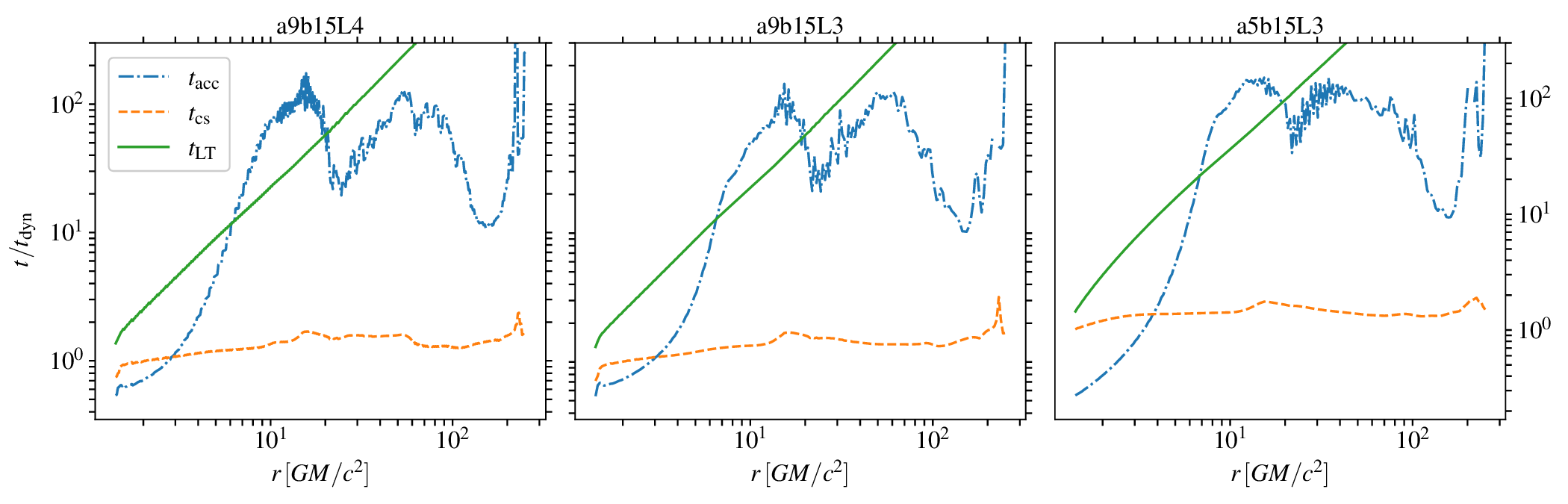}
    \caption{Characteristic timescales (time-averaged over the interval $t\in [10000\,GM/c^3 , t_{\rm end}]$), scaled to the local orbital period, for all three tilted disc simulations. The Lense-Thirring (LT) precession timescales are shorter than the accretion timescales in the thick disc region ($6\lesssim r/r_{\rm g}\lesssim 20$), while the sound-crossing timescales are shorter than the precession timescales everywhere except in the very inner regions of the thick discs, thus allowing for rigid-body precession of the thick disc.}
    \label{fig:timescale}
\end{figure*}

The bottom panel of Fig.~\ref{fig:mdot} shows the mass accretion rate as a function of time at $r=5$ and $25\,GM/c^2$ for all four simulations. Upon comparing the accretion rate at $5\,GM/c^2$ for simulations a9b15L3 and a5b15L3, we find that it is higher for the higher spin case. This could be due to the enhanced angular momentum transport caused by standing-shock features, which become more prominent for higher spins. These standing shocks are characteristic to tilted, thick accretion flows, and are formed along the line of nodes between the black hole symmetry plane and the disc midplane \citep{Fragile08,Generozov14,WQB19}. More notable is that the accretion rates for the misaligned discs exhibit oscillations that stand out clearly in the $r=25\,GM/c^2$ panel of Fig.~\ref{fig:mdot}, while such oscillations are absent for the untilted simulation. We will elaborate on this in Section~\ref{sec:variability}, when we discuss time variability. 

An important diagnostic for tilted accretion discs is comparing characteristic timescales. Fig.~\ref{fig:timescale} illustrates the accretion, $t_\mathrm{acc}=r/\langle V^r\rangle$, sound-crossing, $t_\mathrm{cs}=2 \pi r/\langle c_s \rangle$, and Lense-Thirring precession timescales, $t_{\rm LT} = 2 \pi/\Omega_{\rm LT}$, all normalized by the dynamical timescale, $t_{\rm dyn} = 2 \pi/\langle \Omega \rangle$, for all three tilted simulations. Here, $\langle V^r \rangle$, $\langle c_s \rangle$ and $\langle \Omega \rangle$ are the density-weighted averages of radial velocity, local sound speed and the angular velocity, respectively and the time averaging is done from $t=10000\,GM/c^3$ till the end of the simulation. Also, 
$\Omega_{\rm LT} = \Omega-\Omega_\perp=\Omega\left[1-\sqrt{1-\frac{4a_*}{(r/r_{\rm g})^{3/2}}+\frac{3a_*^2}{(r/r_{\rm g})^2}}\,\right]$ is the Lense-Thirring precession rate, with $\Omega$ the orbital angular velocity, and $\Omega_\perp$ the vertical epicyclic one. For precession to occur, we need the precession timescale to be smaller than the accretion one \citep{BP1975}. Otherwise, the accreted angular momentum can impede precession and lead to disc alignment. Fig.~\ref{fig:timescale} reveals that this criterion is only fulfilled in the thick disc region ($6\lesssim r/r_{\rm g}\lesssim 20$), indicating precession should only be expected in that region. Furthermore, we find that the sound-crossing timescale is shorter than the precession timescale at nearly all radii (excluding the innermost regions), causing a strong coupling of the disc material. This leads us to our anticipation of rigid-body precession \citep{LNPT1996}. We present more details on this in the following section.

\subsection{Tilt \& Twist}
\label{sec:precession}
To investigate the precession and warp in tilted disc simulations, it is useful to evaluate the Euler angles, $\beta$ and $\gamma$, which represent the tilt and twist, respectively. Following the approach introduced in \citet{Fragile07}, we define the tilt angle as the measure of misalignment between the disc and the black hole spin axis as
\begin{equation}
    \beta(r,t) = \cos^{-1} \left[\frac{\bm{J}_{\rm BH} \cdot \bm{J}_{\rm disc}(r,t)}{\abs{\bm{J}_{\rm BH}}    \abs{\bm{J}_{\rm disc}(r,t)}}\right] ~,
\end{equation}
where 
\begin{equation}
    \bm{J}_{\rm BH} = \left(-a_*M^2 \sin \beta_0 \hat{x},\,0,\,a_*M^2 \cos \beta_0 \hat{z}\right)
\end{equation}
is the angular momentum vector of the black hole, and 
\begin{equation}
\left( \bm{J}_{\rm disc} \right)_{\eta} (r,t) =  \frac{\epsilon_{\mu \nu \sigma \eta}L^{\mu \nu}S^{\sigma}}{2\sqrt{-S^{\alpha}S_{\alpha}}}
\end{equation}
for $\eta = 1,2,3$, which correspond to the Cartesian vector components of the disc's angular momentum vector measured in asymptotically flat space. Here,
\begin{equation}
      L^\mathrm{\mu \nu} = \int\left(x^\mu T^\mathrm{\nu 0} - x^\nu T^\mathrm{\mu 0} \right)d^\mathrm{3}x  
\end{equation}
and
\begin{equation}
      S^{\sigma} = \int T^{\sigma 0} d^\mathrm{3}x ~.
\end{equation}
Here $T^{\mu\nu} = \left( \rho \bar{h} +2P_{\rm m}\right)u^{\mu}u^{\nu} + \left(P_{\rm g}+P_{\rm m}\right)g^{\mu \nu}-b^{\mu}b^{\nu}$ is the stress-energy tensor of the fluid in the magneto-hydrodynamic limit, where $\bar{h} = 1+\epsilon+P_{\rm g}/\rho$ is the specific enthalpy, $\epsilon$ is the specific internal energy density, $P_{\rm m}$ is the magnetic pressure, and $b^{\mu}$ is a four-vector magnetic field measured by an observer in the frame co-moving with the fluid, and relates to the divergence-free spatial magnetic field vector through $B^{\mu} = \sqrt{-g}(b^{\mu}u^0-b^0 u^{\mu})$. 

\begin{figure}
\begin{subfigure}[t]{0.47\textwidth}
         \centering
         \includegraphics[width=\textwidth]{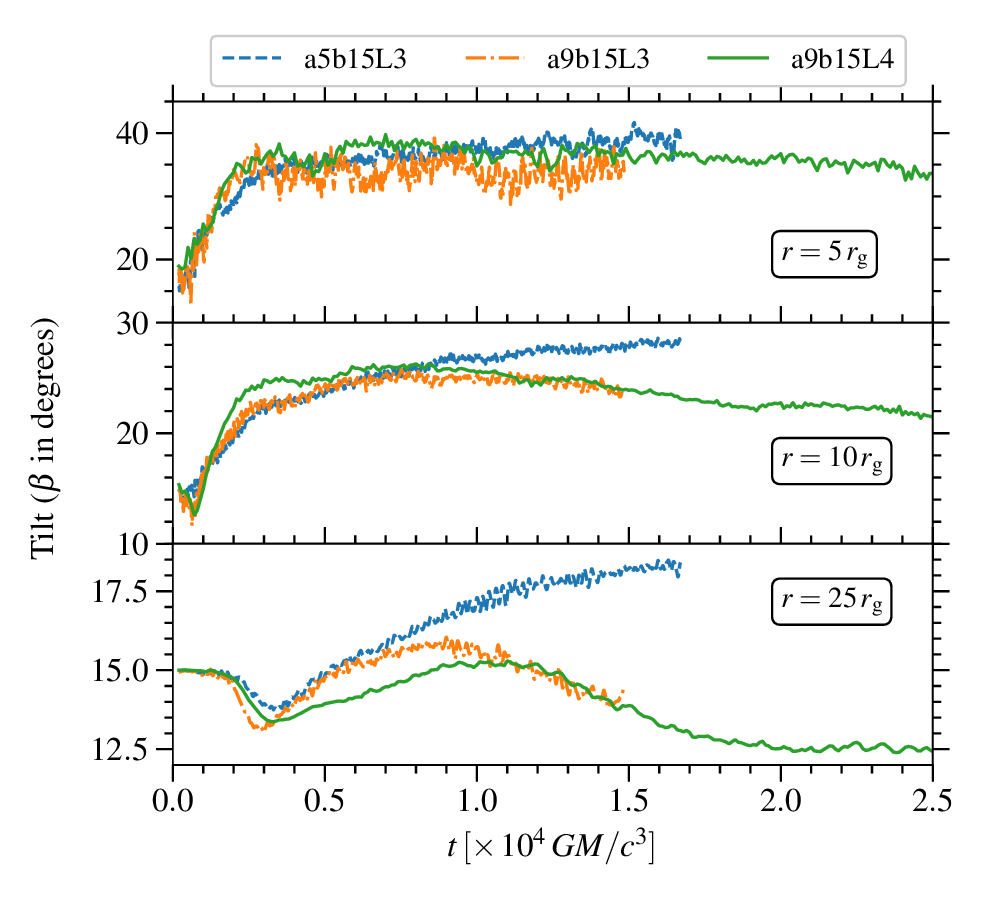}
\end{subfigure}
\caption{Time evolution of the tilt angle, $\beta$, at radii $5$ (top panel), $10$ (middle panel), and $25\,GM/c^2$ (bottom panel) for all three tilted simulations.}
\label{fig:tilt}
\end{figure}

At $t=0$, the angular momentum vector of the disc is aligned along the direction of the $\hat{z}$ unit vector. Fig.~\ref{fig:tilt} shows the evolution of the tilt angle, $\beta$ (measured in degrees), for each of the tilted simulations at three different radii: $5$, $10$, and $25\,GM/c^2$. In all simulations, we find that the thick-disc region closest to the black hole actually initially tilts further away from the black hole before eventually stabilizing at around $35^\circ$ (see the top panel in Fig.~\ref{fig:tilt}). This tendency for tilted thick discs to tilt {\it away} from the black hole symmetry plane at small radii is a common observation in numerical simulations \citep{Fragile07, Liska18, WQB19} and is consistent with expectations in the bending-wave limit \citep{Ivanov97}. At slightly larger radii (middle panel of Fig.~\ref{fig:tilt}), the evolution of the tilt angle exhibits some differences between $a_*=0.5$ and $0.9$ simulations. While $\beta$ initially increases in all simulations, it quickly reaches a stationary value for both the low- and high-resolution $a_*=0.9$ simulations. In contrast, the tilt in the $a_*=0.5$ simulation appears to continue growing over time. We believe that this initial surge in $\beta$ is rather a consequence of the torus responding to the bending waves propagating within it \citep{ZI2011, ZIPF2014}.

\begin{figure}
\begin{subfigure}[t]{0.49\textwidth}
         \centering
         \includegraphics[width=\textwidth]{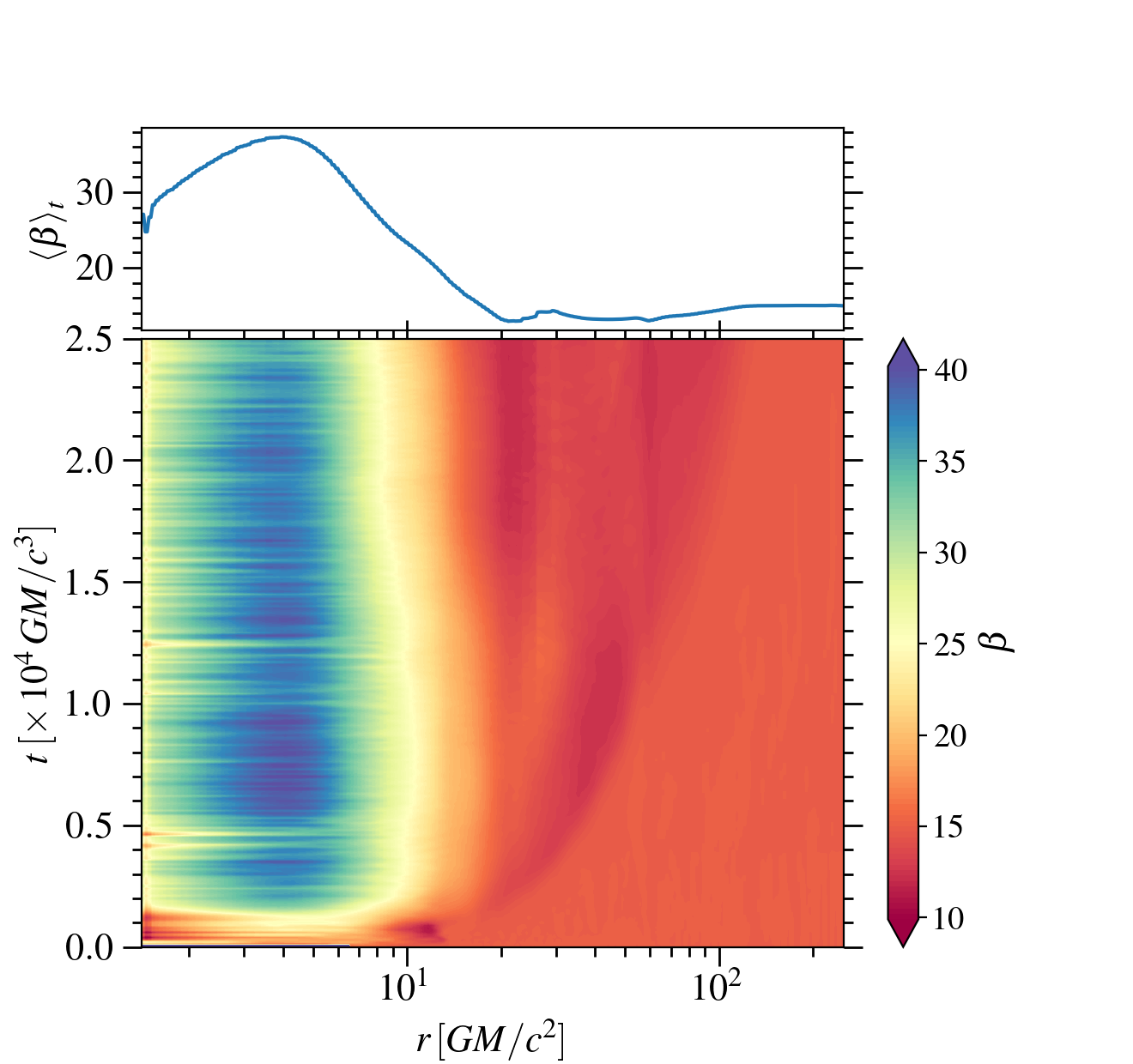}
\end{subfigure}
\caption{Space-time plot for the tilt angle, $\beta$ (measured in degrees), for the a9b15L4 simulation. The blue-solid curve in the top panel shows the radial profile time averaged over $10000-25000\,GM/c^3$.}
\label{fig:tilt2}
\end{figure}

\begin{figure}
\begin{subfigure}[t]{0.49\textwidth}
         \centering
         \includegraphics[width=\textwidth]{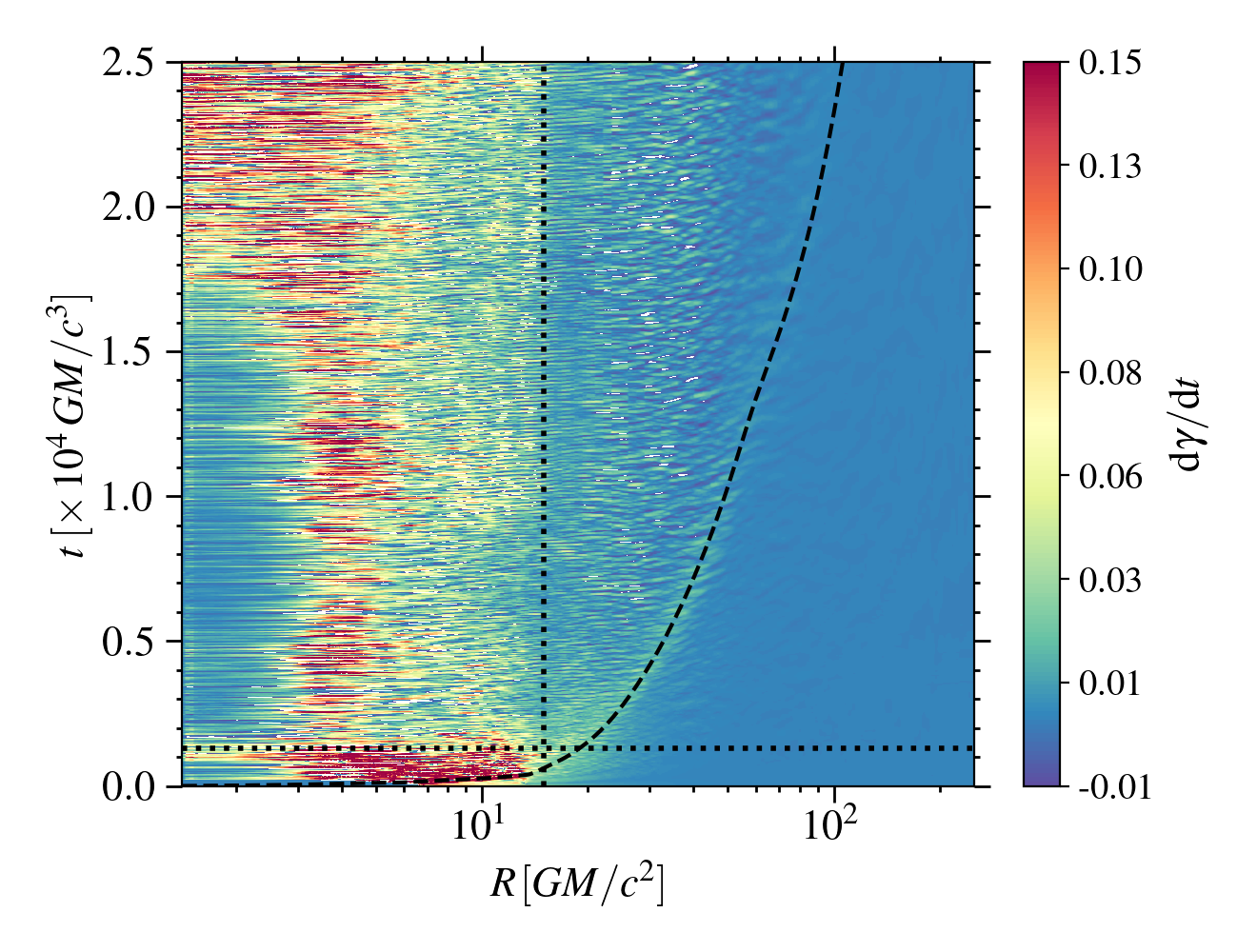}
\end{subfigure}
\caption{Space-time plot for the rate of change of the precession angle. The bending wave, which propagates at roughly half the sound speed, is depicted by the black, dashed curve. The horizontal and the vertical dotted lines mark the time 1300~$GM/c^3$ and the truncation radius 15~$GM/c^2$, respectively. The consistent colour in the precession rate within the range $6-15\,GM/c^2$ suggests rigid-body precession of the thick-disc region.}
\label{fig:prec_rate}
\end{figure}

The decrease in the tilt angle with increasing radius highlights the warped nature of these discs. This can be visualized in Fig.~\ref{fig:tilt2}, where the bottom panel illustrates the time evolution of tilt as a function of radius for simulation a9b15L4. The top panel displays a radial profile of the same quantity, time-averaged over the interval $10000-25000\,\,GM/c^3$.  In the inner regions, within $20\,GM/c^2$, the tilt angle increases as the radius decreases before subsequently decreasing very close to the black hole. This behaviour remains relatively stable over time. The radial profile of $\langle \beta \rangle_{\rm t}$ reveals that the thin disc beyond $20\,GM/c^2$ maintains its initial misalignment at $\beta_0 = 15^{\circ}$, while the inner thick disc and transition region exhibit the expected bending wave pattern \citep{Ivanov97}.

We now calculate the precession angle, $\gamma$, which measures the amount of twisting or precession of the disc's angular momentum vector around the black hole spin axis. The precession angle is calculated as
\begin{eqnarray}
\gamma(r,t) = \cos^{-1}\left[ \frac{\bm{J}_{\rm BH} \times \bm{J}_{\rm disc}(r,t)}{\abs{\bm{J}_{\rm BH} \times \bm{J}_{\rm disc}(r,t)}} \cdot\hat{\textbf{y}}\right] ~.
\label{eq:twist}
\end{eqnarray}
We also consider the projection of $\bm{J}_{\rm BH} \times \bm{J}_{\rm disc}(r)$ onto $\hat{\textbf{x}}$ to avoid the degeneracy in cosine for angles greater than $180^{\circ}$. Fig.~\ref{fig:prec_rate} illustrates the precession rate, $d\gamma/dt$, for the simulation a9b15L4. As previously shown in Fig.~1 of \citet{BFK22}, a precession front originating from the inner thick-disc region propagates outward, causing the disc to warp as the front expands over the bending wave timescale\footnote{Computed as $\int 2{\rm d}r / \langle c_s(r,t)\rangle_{\theta, \phi}$} (depicted by the black dashed curve). The uniform colour in the precession rate within the $6-15\,GM/c^2$ range suggests rigid-body precession\footnote{Our criterion for rigid-body precession is that all annuli at different radii precess at a constant rate, even though each annulus has precessed by a different angle due to an initial, brief period of differential precession.} of the thick-disc region. It is evident from this figure that the precession rate of the inner region  decreases significantly when the bending wave reaches the transition region (close to $15\, GM/c^2$). As we previously reported in \citet{BFK22}, the angular momentum fluxes from the outer thin disc, counteract the Lense-Thirring torque that drives the precession of the inner torus, resulting in a deceleration of its precession.

As the bending wave passes through the thin disc, it also undergoes a brief period of precession. This happens for two reasons: a) the accretion timescales are longer than the precession timescales in the disc before the bending wave reaches it, and b) in all our simulations, the effective stress parameter \footnote{We compute $\alpha$ as the ratio of the $r$-$\phi$ component of stress in the co-moving frame to the total pressure. Both the stress and the pressure terms include contributions from the magnetic field.}, $\alpha$, remains less than the disc scale height everywhere. The latter could be a consequence of under-resolving the outer thin disc, although conducting higher-resolution simulations to test this hypothesis would be computationally expensive. However, once the bending wave has propagated through the thin disc, the accretion timescales become shorter than the precession timescales. Consequently, the thin disc eventually ceases to precess, and any warps are diffused by dynamical processes within the disc. Our simulation of an isolated torus with the same initial setup as the torus in simulation a9b15L4 exhibits the expected precession rate for an isolated torus (see Appendix~\ref{appA} for further details).

\section{Time variability}
\label{sec:variability}
In this section, we perform a comprehensive timing analysis on various accretion flow quantities to investigate the presence of QPOs. To compute the power spectra, we employ the Lomb-Scargle periodogram method \citep{Lomb1976, Scargle1982, V2018}, focusing on density-weighted averages of a quantity, $X$, calculated as follows:
\begin{eqnarray}
    \langle X \rangle_{\theta, \phi}(r,t) = \frac{\int_{\theta}\int_{\phi}\rho(r,\theta,\phi,t) X(r,\theta,\phi,t) \sqrt{-g}{\rm d} \theta {\rm d}\phi}{\int_{\theta}\int_{\phi}\rho(r,\theta,\phi,t) \sqrt{-g}{\rm d} \theta {\rm d}\phi},
\end{eqnarray}
where $\rho(r,\theta,\phi,t)$ represents the density distribution. The power spectra are calculated within a frequency range bounded by the minimum frequency set by the inverse of the time duration being analyzed and by the maximum frequency of $1/(2 \Delta t)$, where $\Delta t$ is the time interval between successive data dumps, which is provided in the last column of Table~\ref{tab:sims}. 

\begin{figure*}
\centering
\includegraphics[width=\textwidth]{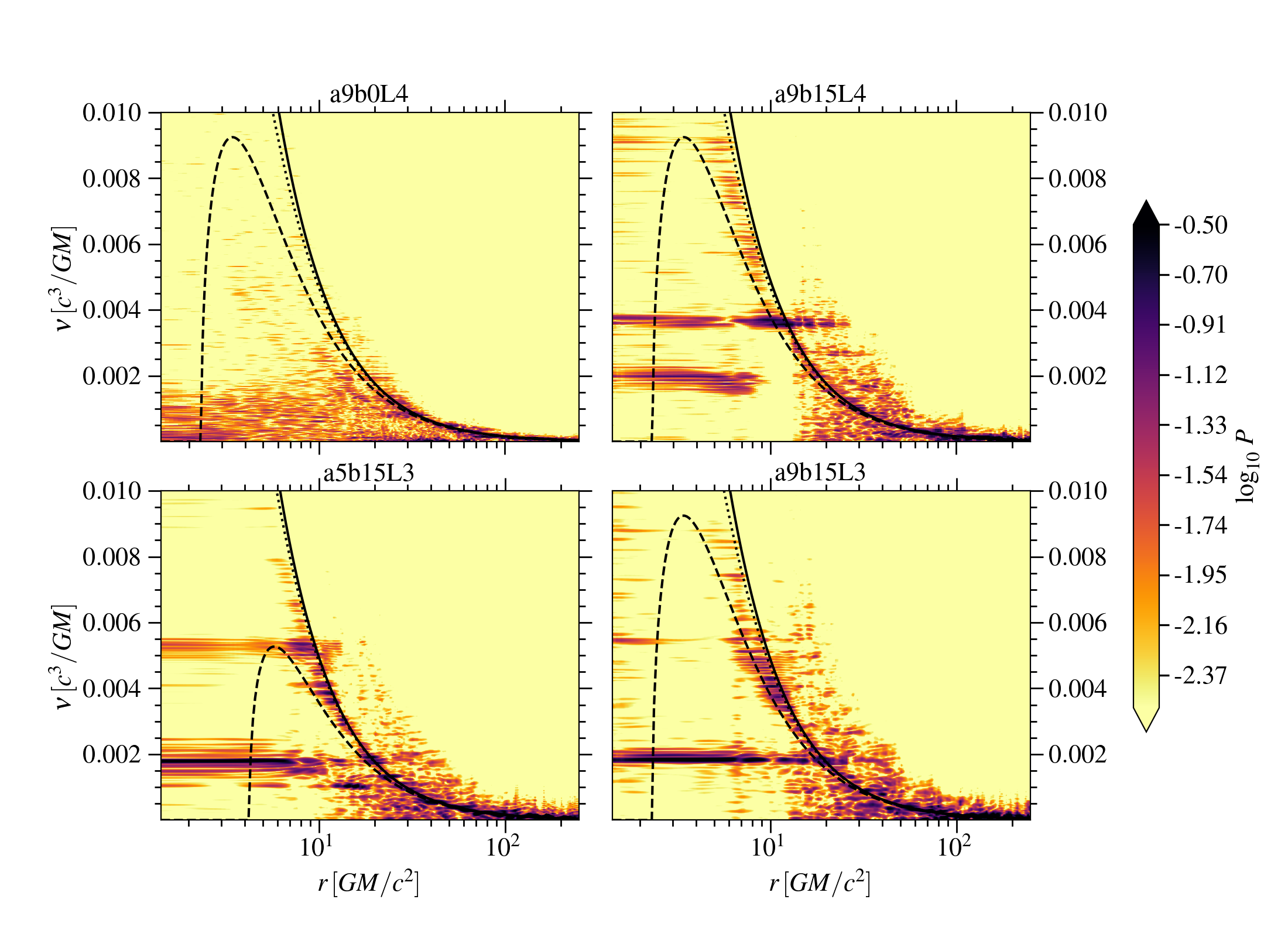}
\caption{Power-spectral densities for all four simulations computed for the density-weighted average of the polar velocity, $V^\theta$. The solid, dashed, and dotted curves correspond to the Keplerian, radial, and vertical epicyclic frequencies, respectively. We note power along or between the epicyclic frequencies in all simulations, as well as horizontal (QPO-like) features in the tilted simulations.}
\label{fig:vthe_psd}
\end{figure*}

Fig.~\ref{fig:vthe_psd} displays the radially-dependent power spectra of the density-weighted average of the poloidal component of the transport velocity\footnote{The transport velocity, $V^i$, is related to the fluid-four velocity through $V^i = u^i/u^t$, for $i=1,2,3$.},$V^{\theta}$, over a spherical shell for all simulations. In the case of the untilted disc (upper-left panel), there is clear power along the epicyclic frequency in the thin disc region; however, there is no evident global QPO-like feature, which would appear as a horizontal band of power. 

In contrast to the untilted disc simulation, the power spectral densities (PSDs) in the tilted simulations exhibit clear signatures of variability at at least two discrete frequencies each, ranging between $0.001$ and $0.006\,c^3/GM$, with all three simulations exhibiting oscillations at a frequency close to $0.002\,c^3/GM$. Given that we are looking at power in the vertical component of velocity, these features suggest that the inner disc undergoes vertical oscillations. We will provide a detailed discussion on the nature of these oscillations in the remainder of this section. 

It is worth noting the enhanced variability power near the vertical epicyclic frequency curve ($\Omega_{\rm z}$, dotted black) or, more accurately, between the vertical and radial epicyclic frequency ($\kappa$, dashed black) curves. Such variability at local epicyclic frequencies is expected in geometrically thin discs \citep{RM2009, MKF2020}, but also in tilted thick discs, as they exhibit eccentric orbital trajectories \citep{FG2008, HBFF2009}.

\begin{figure*}
\centering
\includegraphics[scale=0.55]{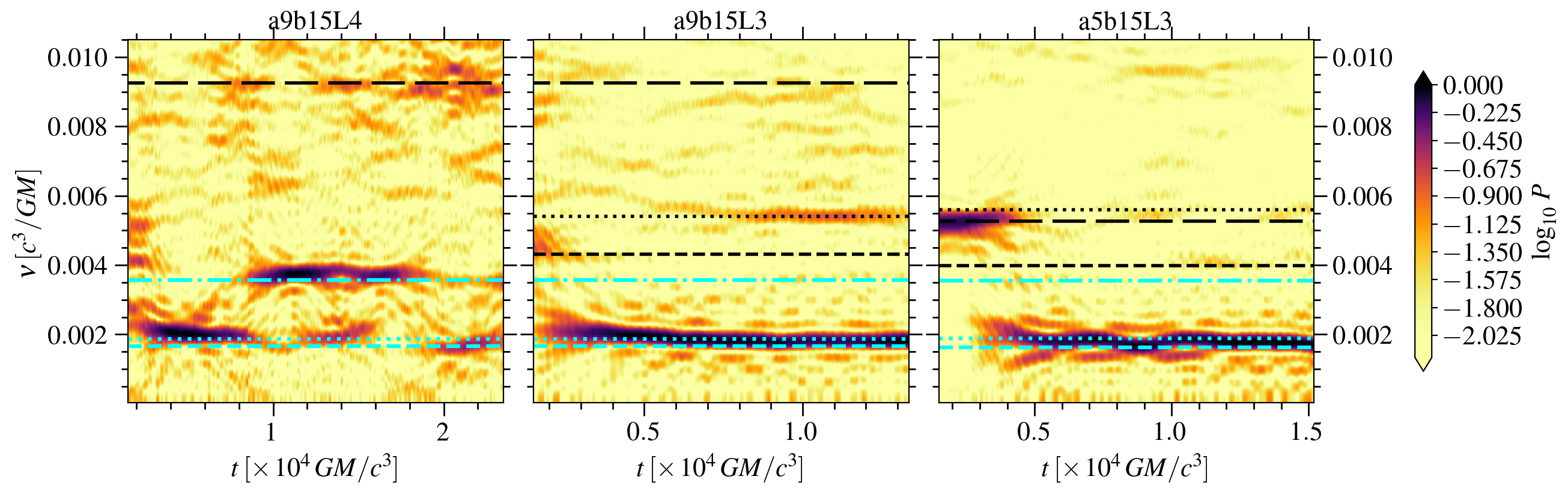}
\caption{Dynamical power spectra of $\langle V^{\theta}\rangle_{\theta, \phi}(r,t)$ at $r=2\,GM/c^2$ for all three tilted disc simulations, computed using a sliding window of length $3000~GM/c^3$. The cyan, dashed and dotted lines correspond to the radial and vertical epicyclic frequencies, respectively, at $19\,GM/c^2$, while the cyan dash-dot line represents the sum of the Keplerian and radial epicyclic frequencies. The long-dashed black lines represent the maximum radial epicyclic frequencies, and the black, dashed and dotted lines in the second and third panel correspond to the radial and vertical epicyclic frequencies, respectively, at the radius of the initial pressure maximum, i.e., $9\,GM/c^2$.}
\label{fig:vthe_dynpsd}
\end{figure*}

\subsection{Trapped $g$-modes}
Trapped $g$-modes\footnote{Note that unless a flow is stably stratified (i.e., has a stable entropy gradient), these potentially trapped ``g"-modes would be inertial modes, rather than gravity modes.} in the inner regions of accretion disc have been studied as a possible source of high-frequency QPOs for a long time \citep{KF80, OKF1987, NoWa1992,FG2008}. While these modes are observed in hydrodynamic simulations \citep{RM2009, DLOF2020I, MKF2020}, they are not commonly observed in magnetohydrodynamic (MHD) discs, possibly due to damping effects associated with the magnetorotational instability (MRI). However, recent studies by \citet{DLOF2020II} have shown that sufficiently strong eccentric distortions can excite trapped inertial waves in MHD simulations. Tilted discs, which exhibit eccentric orbits driven by the tilt \citep{Fragile08}, offer a favourable environment for the survival of these trapped g-modes even in the presence of MRI turbulence. 

In Fig.~\ref{fig:vthe_dynpsd}, we present the dynamical power spectrum of the same quantity, $\langle V^{\theta}\rangle_{\theta, \phi}$, computed at $r=2\,GM/c^2$. While this radius technically falls inside the ISCO for both the 0.5 and 0.9 spin cases, we deliberately chose it to effectively showcase all the QPO features across all three simulations simultaneously. To compute this, we apply a periodogram analysis to a sliding time window of $3000~GM/c^3$. The long-dashed, black line in Fig. \ref{fig:vthe_dynpsd} represents the maximum of the radial epicyclic frequency ($\kappa_{\rm max}$), which is determined by the black hole spin. In the simulations with $a_* = 0.9$, there are some indications of power around $\kappa_{\rm max}$ at later times, although nothing significant compared to the features observed at intermediate frequencies. On the other hand, in the early phase of the $a_* = 0.5$ simulation, there is strong evidence of power along $\kappa_{\rm max}$, though it is challenging to discern whether this power originates from trapped g-modes or is associated with vertical epicyclic motions. The thin-disk dispersion relation (eq. 14 in \citet{FG2008} for the isothermal case) predicts $m=0$ inertial wave propagation where the squared, Doppler-shifted wave frequency is less than the squared, radial epicyclic frequency, i/.e., $\omega^2<\kappa^2$. However, Fig.~\ref{fig:vthe_psd}, particularly for simulation a5b15L3, shows that the power near $\kappa_{\rm max}$ occurs at frequencies greater than the local epicyclic frequency. Therefore, while the trapped g-modes are a possibility in tilted disc simulations, their presence is actively disfavoured by these PSDs. 

\subsection{QPO features}
\label{sec:qpo}
We now shift our focus to the prominent QPO-like features present at intermediate frequencies. The feature at 0.002~$c^3/GM$ (shown in Fig.~\ref{fig:vthe_psd} \& \ref{fig:vthe_dynpsd}), which is present in all the tilted simulations, seems to correspond to the radial/vertical epicyclic frequency near $20\,GM/c^2$. This frequency is associated with the outer edge of the precessing torus, right around where the Lense-Thirring timescale becomes shorter than the accretion timescale (see Fig.~\ref{fig:timescale}). It is intriguing to see that the torus, while precessing as a rigid body within $r\sim 20\,GM/c^2$, also undergoes global epicyclic oscillations at a frequency corresponding to its outer edge. The cyan dashed and dotted horizontal lines around 0.002~$c^3/GM$ in Fig.~\ref{fig:vthe_dynpsd} represent the radial and vertical epicyclic frequencies at $19\,GM/c^2$, respectively. 

The relatively high-frequency feature in these simulations is more complex. In the low-resolution simulations, the feature at around $0.005\,c^3/GM$ matches with the vertical epicyclic frequency at the location of the initial pressure maximum of the torus, which is $9\,GM/c^2$. This frequency is shown by the black, dotted horizontal line in Fig.~\ref{fig:vthe_dynpsd}. However, this correspondence does not seem to hold for the high-resolution simulation. In a9b15L4, the higher frequency feature is instead observed around 0.004~$c^3/GM$, which coincides with the sum of the Keplerian and radial epicyclic frequencies at $19\,GM/c^2$ (indicated by the cyan dash-dot line). Our current understanding is that the mesh refinement at $10.4\,GM/c^2$ may be influencing this feature, potentially enhancing it over the epicyclic frequencies at $9\,GM/c^2$, given that the local vertical epicyclic frequency at the mesh refinement radius happens to also be close to $0.004~c^3/GM$.

In the literature, tori examined for the study of various oscillation modes are typically isolated and exhibit polytropic behavior. The oscillation frequencies of such tori are closely tied to the characteristic frequencies at their pressure maxima. This connection arises because even if each annulus of the torus oscillates at its local epicyclic frequency, the average oscillation frequency converges toward the epicyclic frequency of the pressure maximum, given that most of the torus's mass is concentrated there. Consequently, it is unsurprising to observe that the feature at $0.002\,c^3/GM$ corresponds to the radial/vertical epicyclic frequency at approximately $19\,GM/c^2$, as that is where the bulk of the mass resides in our simulations. Although we initialized the torus with the pressure and density maximum at $9\,GM/c^2$, the presence of the outer thin disc, with a density at least an order of magnitude higher, causes the pressure maximum and density maximum of the precessing torus to migrate towards its outer edge (see Fig.~\ref{fig:sigma}). Additionally, the strong coupling of the thick disc within $20\,GM/c^2$, induced by the bending waves due to the Lense-Thirring effect, further encourages global oscillations. It is thus intriguing to observe the profound influence of the outer thin disc on the inner torus, as it modifies its shape and mass distribution, and consequently, promotes eigenfrequencies that diverge significantly from those of an isolated torus.

Furthermore, we have observed that the Brunt-Väisälä frequency exceeds the local epicyclic frequencies within the torus region but is comparable to the local epicyclic frequencies in the outer thin disc region. Therefore, it is plausible that the oscillations in our simulations are influenced to some extent by a buoyancy restoring force. While this may not significantly alter the types of nonlinear interactions that could excite oscillations, it may complicate comparisons with oscillations in isolated tori, as seen in previous studies \citep[e.g.,][]{Blaes06, Mishra17}.

\subsection{Radial and Vertical epicyclic motions}
\label{sec:epicyclic}
The presence of power over a range of radii at a fixed frequency suggests that the torus is undergoing the motions of some global mode. Earlier studies \citep[e.g.,][]{Fragile08} reported global radial epicyclic motions in tilted disc simulations, but those motions were not correlated with any particular QPO frequency. To further investigate the nature of these oscillations, we compute the PSD of the density-weighted average velocities for different modes, represented as $\langle \rho V^{i} \cos(m\phi)\sin(n\theta-\theta_{\rm mid})\rangle_{\theta,\phi}/ \langle \rho \rangle_{\theta,\phi}$, for $i=r$ and $\theta$. Here, $m$ represents the azimuthal wavenumber and $n$ denotes the number of nodes in the vertical direction \footnote{Note that oscillations' structure in $\theta$ may not be described by a single Fourier mode $\sin(n\theta-\theta_{\rm mid})$, but using this projection allows us to distinguish between modes with/without vertical structure.}. For this analysis, we compute the eigenfunctions in the transformed coordinates, where the midplane of the disc is aligned with the $x^\prime$-$y^\prime$ coordinate plane so that $\theta_{\rm mid}^\prime = \pi/2$. The results remain the same if we instead repeat the analysis in the coordinate frame using $\theta_{\rm mid}$ from Eq.~\ref{eq:thetamid}.

\begin{figure*}
\centering
\includegraphics[width=\textwidth]{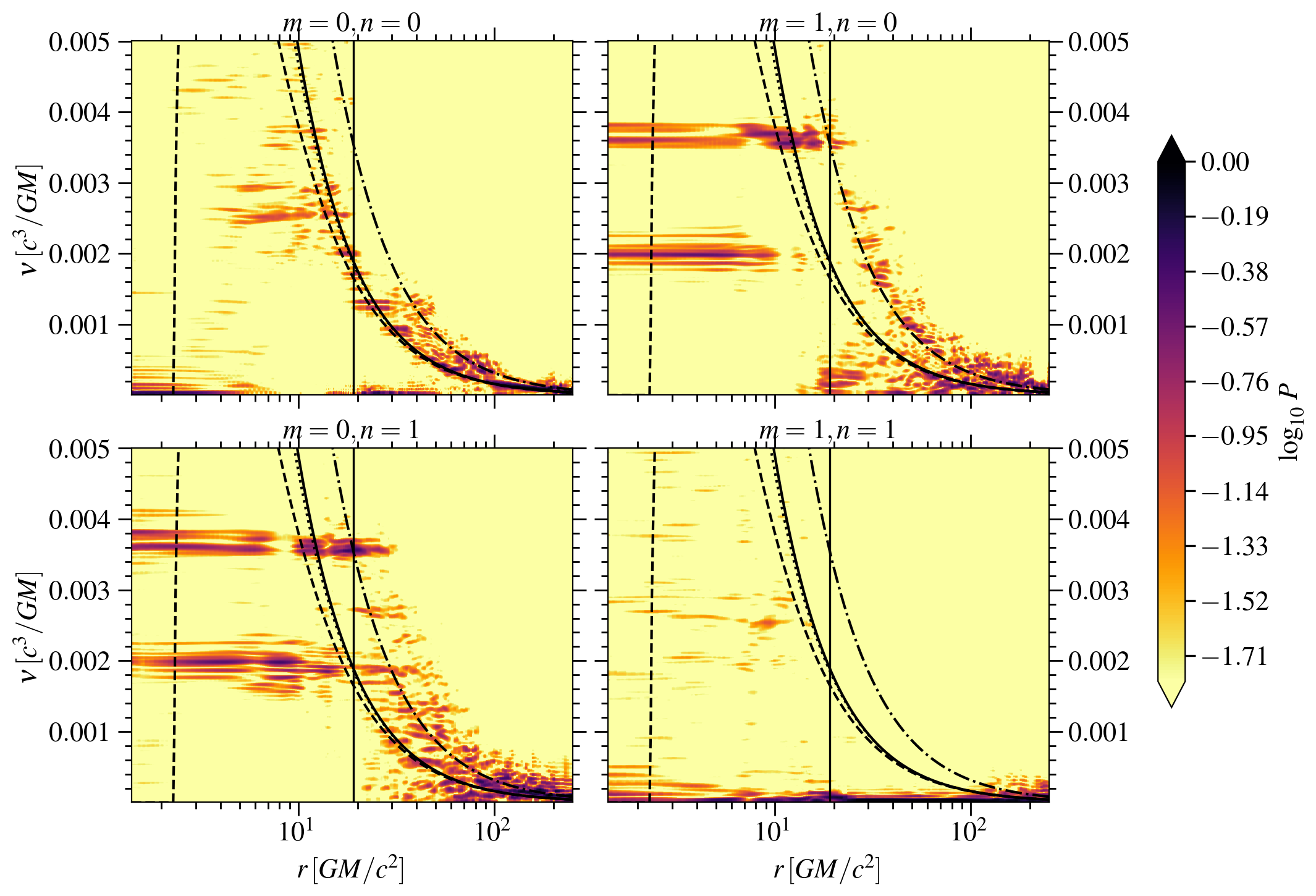}
\caption{Power-spectral densities of $\langle \rho V^{r^\prime} \cos(m\phi^\prime)\sin(n\theta^\prime-\theta_{\rm mid}^\prime)\rangle_{\theta^\prime,\phi^\prime}/ \langle \rho \rangle_{\theta^\prime,\phi^\prime}$ for all combinations of $m$, $n=0$ and 1 for the high-resolution, tilted simulation, a9b15L4. The solid, dashed and dotted curves correspond to the Keplerian, radial and vertical epicyclic frequencies, respectively. The dashdot curve represents $m\Omega_{\rm K}+\kappa$, while the vertical solid line marks $19\,GM/c^2$.}
\label{fig:vr_modepsd}
\end{figure*}

Fig.~\ref{fig:vr_modepsd} displays the PSD of $V^{r^\prime}$ for all combinations of $m$ and $n$ between 0 and 1. The overlaid curves carry the same meaning as in Fig.~\ref{fig:vthe_psd}. In addition to these curves, we include a dash-dot curve representing $m\Omega_{\rm K}+\kappa$ and a vertical solid line at $19\,GM/c^2$ as a reference point. Here $\Omega_{\rm K}$ and $\kappa$ represent the Keplerian and radial epicyclic frequency, respectively. We find power in the $m=1$, $n=0$ panel that resembles radial epicyclic motion, e.g., the entire torus oscillating around the black hole radially. The power in $m=0$, $n=1$ maybe associated with the radial pulsations with a node at the disc midplane, where one half of the disc is radially expanding while the other half is radially contracting. Additionally, there is power along the $m\Omega_K+\kappa$ curves in the outer, thin-disc region, indicating that while the thin disc oscillates at the local epicyclic frequency, the thick disc undergoes global oscillations.

\begin{figure*}
\centering
\includegraphics[width=\textwidth]{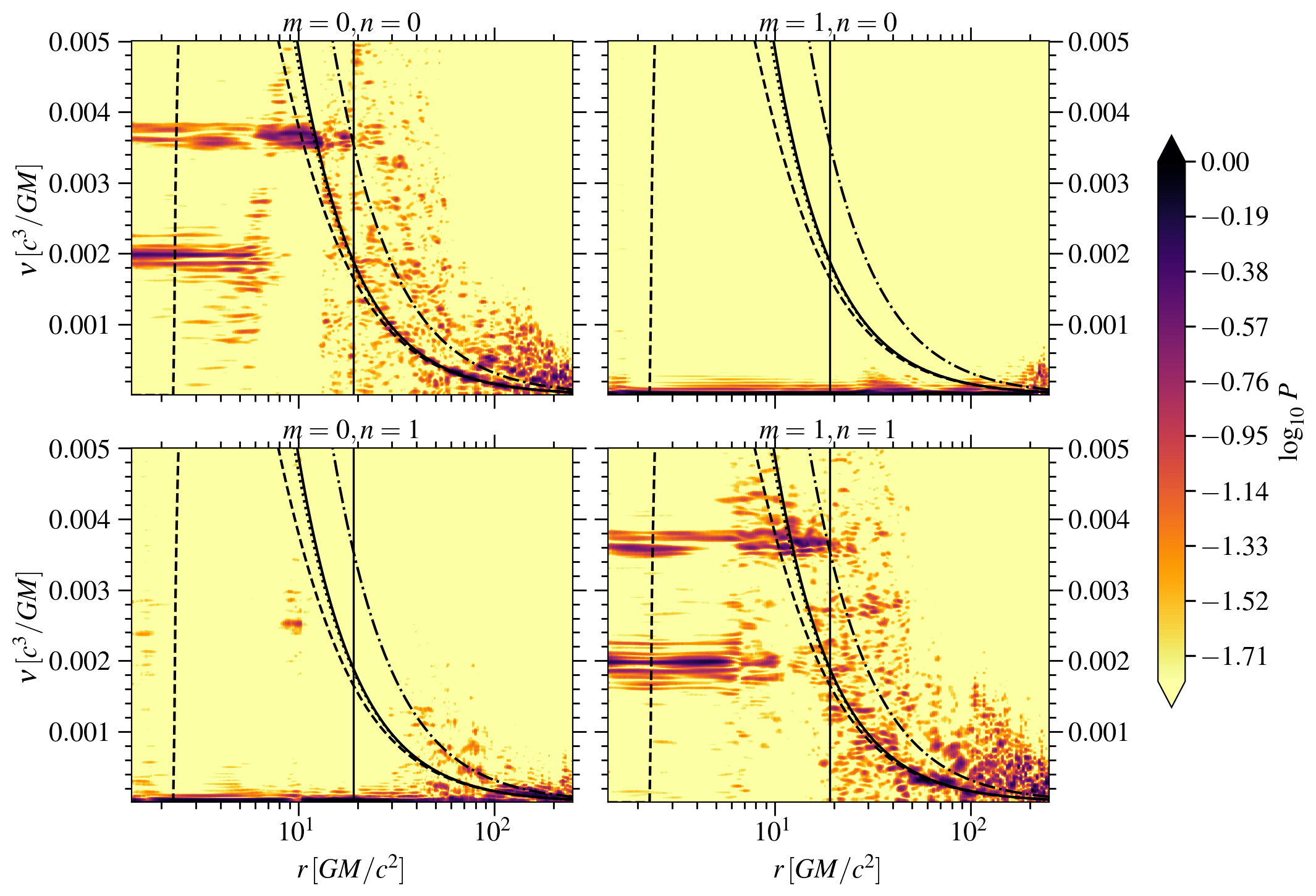}
\caption{Power-spectral densities of $\langle \rho V^{\theta^\prime} \cos(m\phi)\sin(n\theta-\theta_{\rm mid})\rangle_{\theta^\prime,\phi^\prime}/ \langle \rho \rangle_{\theta^\prime,\phi^\prime}$ for all combinations of $m,n=0$ and 1 for the high-resolution, tilted simulation, a9b15L4. The lines and curves carry the same meaning as in Fig.~\ref{fig:vr_modepsd}.}
\label{fig:vth_modepsd}
\end{figure*}

Fig~\ref{fig:vth_modepsd} shows the same analysis as in Fig.~\ref{fig:vr_modepsd}, but for $V^{\theta^\prime}$. Note that we now observe power in the opposite panels, namely $m=0$, $n=0$ and $m=1$, $n=1$. The $m=0$, $n=0$ mode indicates pure vertical oscillations in the disc, while the $m=1$, $n=1$ mode corresponds to a motion resembling non-axisymmetric, vertically pulsating mode. The power spectra, including mode numbers $>1$, reveal that combinations with odd differences between the mode numbers show power in $V^{r^\prime}$, while combinations with even differences between $m$ and $n$ exhibit power in $V^{\theta^\prime}$. The oscillations in $V^{r^\prime}$ and $V^{\theta^\prime}$ are also prominently evident in the respective space-time diagrams, as shown in the Appendix (see Fig.~\ref{fig:vrvth}).

The excitation mechanism for these oscillations and the nature of these modes are not entirely clear. However, previous analytical calculations \citep{Kato2004, kato2008, FG2008} and subsequent simulation studies \citep{HBFF2009, DLOF2020II} have shown that disc warps and eccentricities can excite and amplify inertial modes in the disc through non-linear coupling. These excitation mechanisms can be described as nonlinear three-wave couplings between a nearly zero-frequency distortion with $m\sim 1$, and a pair of waves with nearly equal frequencies and $m$ values differing by one. The appearance of both $m=0$ and $m=1$ power at the same frequencies in Figs. \ref{fig:vr_modepsd} and \ref{fig:vth_modepsd} indicates that such a parametric-type instability may be responsible for the oscillations in our simulations. 

\subsection{Vertical oscillations}
\begin{figure}
\begin{subfigure}[t]{0.495\textwidth}
         \centering
         \includegraphics[width=\textwidth]{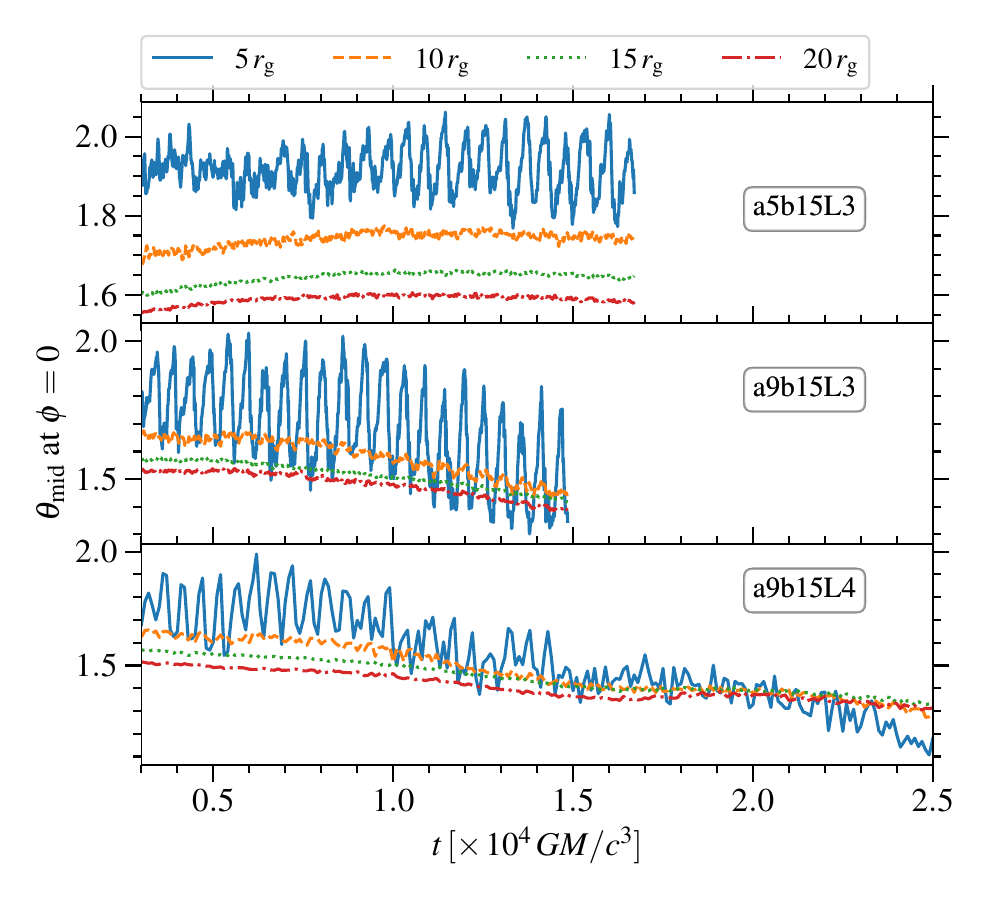}
\end{subfigure}
\begin{subfigure}[t]{0.495\textwidth}
         \centering
         \includegraphics[scale=0.55]{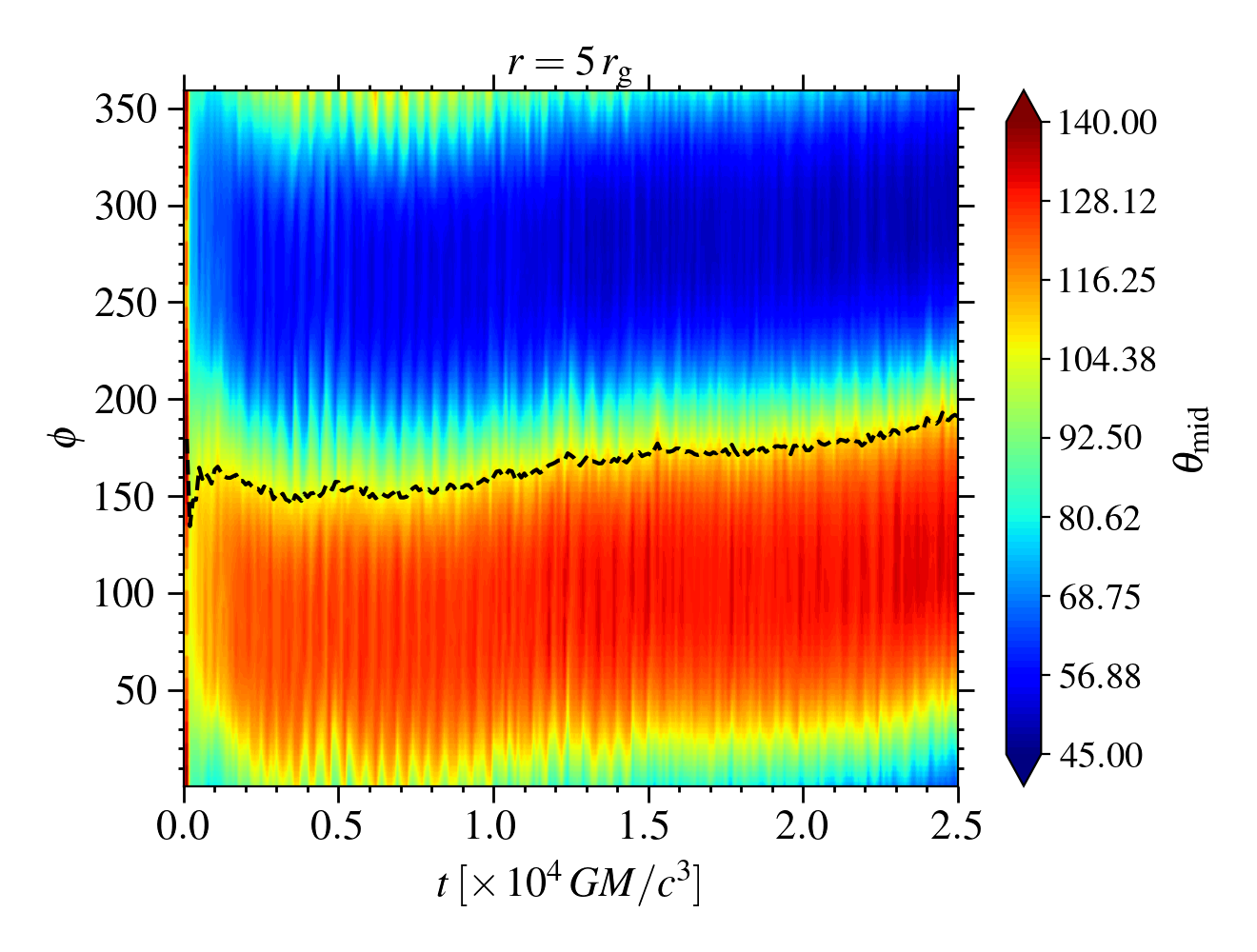}
\end{subfigure}
\caption{Top: Time evolution of the disc's midplane angle, $\theta_\mathrm{mid}$, measured at a fixed azimuthal angle, $\phi=0$, for radii $r=5,10,15$, and $20\,GM/c^2$ for the three tilted-disc simulations. Bottom: A time-space plot of the midplane angle, $\theta_\mathrm{mid}$, at $r=5\,GM/c^2$ for the simulation a9b15L4. A clear $m=1$ pattern is seen which precesses along with the rest of the thick-disc region (black, dashed line) and also oscillates with a period close to $500 r_{\rm g}/c$.}
\label{fig:thetamid}
\end{figure}

Even without the aid of a PSD, the vertical oscillations reported above can be clearly seen in a plot that tracks the disc midplane angle, $\theta_\mathrm{mid}$, over time, as we show in the top panel of Fig~\ref{fig:thetamid}. We compute $\theta_\mathrm{mid}$ based on the concentration of the disc's mass:
\begin{eqnarray}
\theta_{\rm mid}(r,\phi,t) =  \left( \frac{\int_{0}^{\pi}  \rho^2 \theta^2 \sqrt{-g} \ d\theta}{\int_{0}^{\pi} \ \rho^2 \sqrt{-g} \ d\theta} \right)^{1/2} .
\label{eq:thetamid}
\end{eqnarray}
For clarity, we show the variations of $\theta_{\rm mid}$ over time for a specific azimuthal angle ($\phi=0$) at different radii. At $r=5\,GM/c^2$, clear oscillations in $\theta_{\rm mid}$ can be observed in each case, indicating an up-and-down motion of the disc. The amplitude of these oscillations decreases as one moves away from the black hole.

The bottom panel of Fig. \ref{fig:thetamid} provides insight into the azimuthal distribution and evolution of these vertical oscillations over time. A dominant $m=1$ pattern is evident, arising from the deviation of the disc midplane from the grid plane $z=0$. Although the disc starts with $\theta_{\rm mid}=\pi/2$ for all $\phi$, once the precession begins, half of the disc has $\theta_{\rm mid} < \pi/2$ while the other half has $\theta_{\rm mid} > \pi/2$. The $m=1$ pattern shifts with time due to the precession of the thick disc. The black, dashed curve in Fig.~\ref{fig:thetamid} corresponds to the precession rate evaluated at $5\,GM/c^2$ obtained directly from the simulation data. In addition to the evidence this plot provides for tilt and precession, we also observe additional regular and moderately rapid, oscillations of the maximum and minimum values of $\theta_{\rm mid}$, corresponding to the oscillations seen in the top panel. The period of these oscillations is approximately $500\,GM/c^3$, matching the frequency of $0.002\,c^3/GM$ noted in the PSD of polar velocity (Fig. \ref{fig:vthe_psd}). In other words, we are seeing the global $m=0$, $n=0$ vertical oscillations discussed earlier.

\section{Discussion}
\label{sec:discussion}
\subsection{Comparison with previous works}
In this paper, we have investigated the timing and dynamical properties of tilted, truncated accretion discs. While the truncated disc picture is rather strongly motivated by observations of the hard state in X-ray binary systems, their formation from first principles is not yet fully understood. Recent simulations have made progress in self-consistently forming truncated discs by incorporating strong magnetic fluxes. In these simulations, a thin disc initiated with strong magnetic flux results in the accreting gas dragging in enough poloidal magnetic flux to saturate the black hole and disrupt the accretion flow. This leads to the formation of a thin disc that is truncated beyond the normal marginally stable radius, with a hot, magnetically saturated corona present inside \citep{MFAB2022, LMTPB2022, DVF2022}. Another proposed mechanism for truncated disc formation, known as the `disc evaporation model' \citep{MM1994, LYMM1999}, suggests that the inner regions of the disc are evaporated at low densities. However, to further investigate phenomena like disc evaporation, additional physics such as thermal conduction needs to be incorporated into simulations, which would be both challenging and computationally expensive. 

Numerical simulations conducted thus far suggest that strongly magnetized discs do not undergo precession \citep{MTB2013, RWQ2023,Fragile23}. This is attributed to the magnetic torque exerted by the strong fields accumulated near the black hole, which effectively aligns the disc with the symmetry plane of the black hole. Since our primary motive is to study precession in a truncated disc, we chose a magnetic field configuration that leads to a weakly magnetized corona in the inner regions \citep[like a Standard and Normal Evolution (SANE) accretion flow][]{NSPK2012} and maintain the outer, truncated thin disc with the aid of artificial cooling. In the future, we wish to explore the precession of strongly magnetized discs with a truncated disc geometry.

Geometrically thin discs with extreme tilt angles in the presence of a spinning black hole can experience disc tearing, as highlighted in previous studies \citep{NK2012, RAD2021, LHTI2021}. Notably, \citet{MLPVI2023} conducted further analysis of the tearing disc simulations presented in \citet{LHTI2021} and identified high-frequency quasi-periodic oscillation (HFQPO) features corresponding to the radial epicyclic frequencies at the tearing radius. However, it is crucial to emphasize that our simulations differ significantly from the disc tearing scenario. In our simulations, there is no occurrence of disc tearing, and the torus exhibits global oscillations across a range of radii.

Oscillations at epicyclic frequencies have previously been reported in previous global relativistic hydrodynamic simulations of pressure-supported (non-accreting) tori around black holes \citep{SC2006, Mishra17}. Yet, in GRMHD simulations of accreting tori, there has been a lack of compelling evidence supporting the existence of QPOs associated with these oscillations \citep{SKH2006}. Conversely, GRMHD simulations of tilted accreting tori initially reported global variability at specific orbital frequencies \citep{HBFF2009}, only to later identify this as a transient feature \citep{HBF2012}. It was also suggested that the presence of shocks in tilted discs enhances this variability, thus explaining the limited variability seen in aligned, accreting tori. Our results agree with the latter aspect of these studies, as we do not observe any variability in the aligned case. However, in contrast to the earlier studies on tilted accreting tori \citep{HBFF2009, HBF2012}, we have identified a persistent QPO feature at $0.002,c^3/GM$ in all our tilted disc simulations\footnote{Note that even our low-resolution simulations are run for a longer duration than theirs, roughly twice the duration of most of their simulations.}. This feature appears to correspond to epicyclic frequencies at the outer edge of the precessing torus, influenced by the presence of the outer thin disc -- a component absent in any previous simulations documented in the literature.

Our findings contradict the conclusions of \citet{MN21}, who argued that hot, thick accretion flows must be supersonic, and therefore cannot precess as solid bodies. Our results, as depicted in Fig. \ref{fig:timescale}, demonstrate that our accretion discs are not supersonic, even in the hot, thick regions. In fact, $V^r < c_s$ except for the very innermost regions. Furthermore, both in our previous work \citep{BFK22} and here (see Fig.~\ref{fig:prec_rate}), we demonstrated that the hot, thick disc precesses as a rigid body, albeit at a rate slower than predicted for an isolated disc, owing to the presence of the outer thin disc. The only question that remains is whether our simulations can reproduce the spectral and timing properties of the luminous, hard state, particularly the prominent type-C QPO. We plan to address this question in future investigations.

\subsection{Observational relevance}
Often, models attempting to explain QPOs as being associated with the epicyclic frequencies in geometrically thin discs face challenges in matching the observed frequencies due to the radial dependence of these frequencies. For example, HFQPOs that are commonly found in 3:2 frequency ratio pairs \citep{Remillard06} are suggested to be due to a resonance \citep{KA2001a, AK2001}, possibly a parametric resonance \citep{KA2002} of oscillations at the radial and vertical epicyclic frequencies. In contrast, global oscillations in geometrically thick discs are also a viable explanation, as suggested by various studies \citep{RYZ2003, SC2006, Blaes06,Mishra17}. Our findings, as discussed in Sections~\ref{sec:qpo} and \ref{sec:epicyclic}, provide compelling evidence that global oscillations can only be excited in the thick disc occupying the inner region of truncated discs, and only when the discs are misaligned. Thus, our results strengthen the argument that the presence of tilt is a crucial requirement for the occurrence of QPOs.

In addition to the low-frequency QPO, the PSD of black hole transients reveals a more complex structure, typically analyzed using a combination of Lorentzian components \citep{2002ApJ...572..392B}. These components yield characteristic frequencies, but due to their broad nature, determining a single characteristic frequency, as is the case with a narrow peak, becomes a model-dependent task. During the early phases of the state transition, it is observed that as the source flux increases, these characteristic frequencies increase \citep{BL2014, Bhargava2021}. This increase is often attributed to the inward movement of the truncation radius. Our findings, which associate the QPO feature at $\nu \approx 0.002\,c^3/GM$ with epicyclic frequencies at the transition radius (i.e., the outer edge of the precessing torus), are in line with this understanding, since the epicyclic frequency increases with decreasing radius. To further validate this connection, we intend to conduct future truncated disc simulations with different torus sizes, representing varying transition radii, to examine their impact on QPO frequencies. 

As per the findings of \citet{FSB16}, the correlation between the QPO frequencies observed in low mass X-ray binaries could be explained by associating the type-C QPO with the torus precession (equivalent to the $m=-1$ vertical epicyclic mode), and the HFQPOs, often occurring in a near frequency 3:2 ratio, with the axisymmetric modes of the torus, specifically the breathing and vertical epicyclic modes, respectively. The current results from our simulations support this model, where we clearly see a $m=0$ vertically epicyclic mode for the lower high frequency QPO in addition to the precession of the torus, which could be interpreted as an $m=-1$ vertical epicyclic mode. Since our simulations do not cover a full precession cycle, it is not possible to observe any such mode in our PSDs. We note that the other HFQPO we found in our simulations is not a pure breathing mode as in the \citet{FSB16} model, but rather something similar to a non-axisymmetric, vertically pulsating mode. 

Another point is that HFQPOs are only observed in high-luminosity states at specific hardness ratios \citep{MR2006, BSM12}. While this could be a selection effect due to the low signal-to-noise ratio in the low-luminosity states, it could also be that QPOs produced from global oscillations only have a detectable quality factor for small sizes of tori, reached as the source transits to the soft state. 

Further, recent multiwavelength observations of black hole X-ray binaries have enabled detection of LFQPOs both in IR/Optical and very high energies$(>\,200\,\mathrm{keV})$, at nearly the same frequency as the type-C QPO observed in X-rays \citep[e.g.][]{Kalamkar2016, jetnature2021}. The popular explanation for such multi-wavelength observation of a common QPO frequency is a simultaneous precession of the jet along with the corona \citep{Malzac18}. Additional correlated QPO frequencies may be observed in the future if, for instance, the vertical oscillations observed in the hot, thick inner disc of our simulations were to propagate into the jet, which may happen if this region of the disc plays a role in confining or collimating the base of the jet.

\section{Summary}
\label{sec:conclusions}
In this paper, we performed a suite of GRMHD simulations of truncated accretion discs, tilted and untilted, to study the dynamics and precession of the hot, inner, geometrically thick flow in the presence of a cold, outer, geometrically thin disc. Our main conclusions are:
\begin{itemize}
    \item {\it There is a slow down of the precession rate}: The precession rate of the inner torus in a truncated geometry is significantly slower than the precession rate of an isolated torus of the same size, as we noted previously in \citet{BFK22}. This decrease in the precession rate is caused by the flux of angular momentum between the two disc components.
    \item {\it Tilted discs exhibit significantly more variability}:  We observe QPO-like features at different frequencies in all of our tilted disc simulations, whereas such features are absent in the untilted case.
   
    \item {\it The inner, misaligned torus exhibits coherent vertical oscillations and radial epicyclic motions}: We identify a prominent QPO-like feature at a frequency of approximately $0.002\,c^3/GM$, which is consistently present in the PSDs of all three tilted simulations. This frequency corresponds to approximately 40 Hz for a black hole of $10\,M_{\odot}$. This feature seems to correspond to the radial or vertical epicyclic frequency at the transition radius.

    We also observe another QPO-like feature at approximately $0.004\,c^3/GM$ in the high-resolution simulation, a9b15L4, and approximately $0.005\,c^3/GM$ in the low-resolution simulations, a9b15L3 and a5b15L3; in the a5b15L3 case its frequency could be in a 3:1 ratio to the fundamental. We hypothesize that the feature in the high-resolution simulation corresponds to the vertical epicyclic frequency at the mesh refinement radius, while the feature in the low-resolution simulations corresponds to the vertical epicyclic frequency at the radius of the initial pressure maximum in the torus.

    \item {\it The thin disc exhibits variability at the local epicyclic frequency corresponding to each radius}: In addition to the QPO-like features associated with the radial and vertical epicyclic motions of the inner thick disc, our simulations reveal power along the Keplerian frequency curve in the thin disc region, indicating local oscillations at the corresponding Keplerian frequencies. 
    
    Overall, our results align well with the observations of HFQPOs in the black hole X-ray binary systems. 
\end{itemize}

\section*{Acknowledgements}
We express our gratitude to the referee, Pavel Ivanov, for his meticulous comments and valuable suggestions on the manuscript. We would also like to thank Omer Blaes and Earl Bellinger for helpful discussions regarding this work. Some of the simulations were performed on the Prometheus cluster, part of the PL-Grid infrastructure located at ACK Cyfronet AGH in Poland and the Max Planck Computing \& Data Facility. This work also used the Extreme Science and Engineering Discovery Environment (XSEDE), which is supported by National Science Foundation grant number ACI-1548562. PCF gratefully acknowledges the support of the National Science Foundation through grants AST1616185, PHY-1748958, and AST-1907850. This work was performed in part at the Aspen Center for Physics, which is supported by National Science Foundation grant PHY-1607611. 
\section*{Data Availability}
The data underlying this paper will be shared upon reasonable request to the corresponding author. 
\bibliographystyle{mnras}
\bibliography{thinThick.bib} 

\appendix
\section{Simulation of an isolated torus}
\label{appA}
\begin{figure}
\begin{subfigure}[t]{0.47\textwidth}
         \centering
         \includegraphics[width=\textwidth]{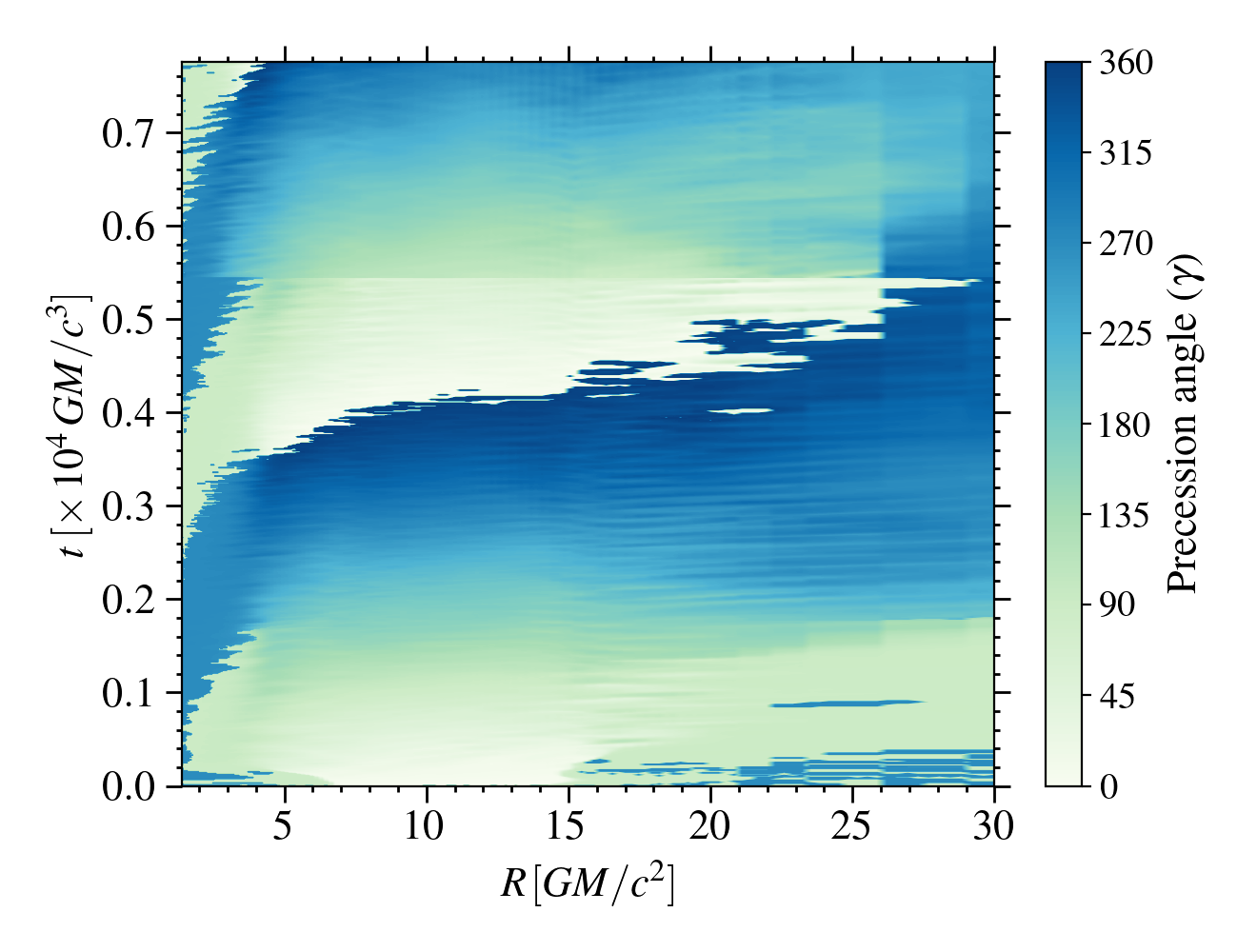}
\end{subfigure}
\caption{Space-time evolution of the precession angle, $\gamma$ (measured in degrees), for an isolated torus simulation.}
\label{fig:prec_torus}
\end{figure}

To ensure that the slowdown in the precession rate is a physical effect caused by the outer thin disc rather than a numerical artifact, we performed one additional simulation of an isolated torus using the exact same numerical setup as in simulation a9b15L4, but excluding the thin disc beyond $15\,GM/c^2$. Fig~\ref{fig:prec_torus} shows the time evolution of the precession angle. We find that the torus completes one full precession cycle within $4500~GM/c^3$. This is consistent with the expected precession rate for an isolated torus extending from 5 to $15\,GM/c^2$ \citep[see eq. 2 in][]{Ingram09}, but much faster than what we see when the thin disc is included.

\section{Supplementary Figures}
\label{appB}
\begin{figure}
\begin{subfigure}[t]{0.47\textwidth}
         \centering
         \includegraphics[width=\textwidth]{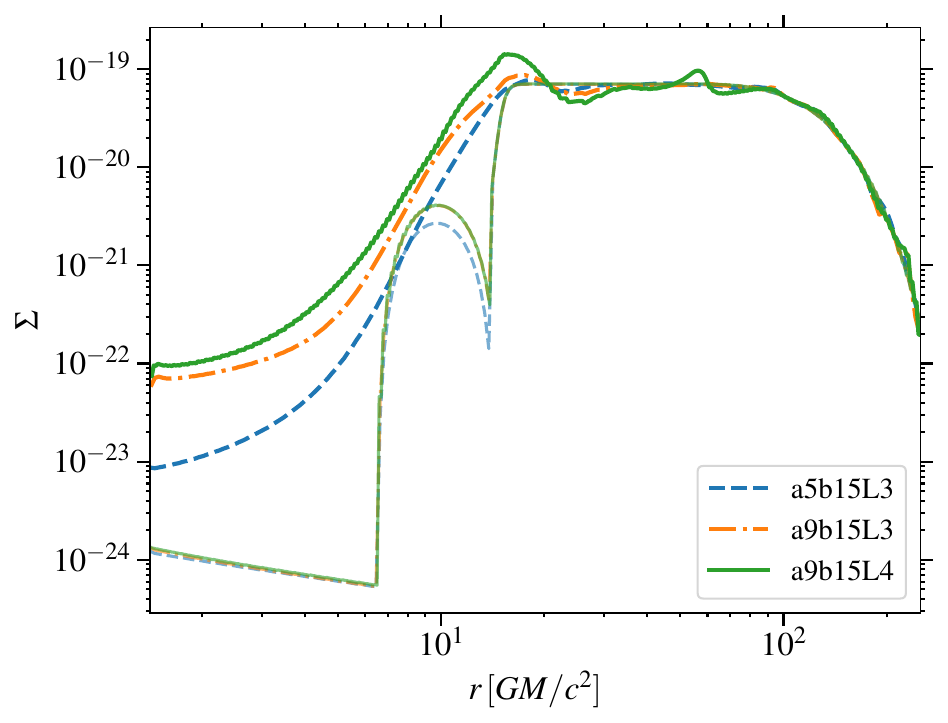}
\end{subfigure}
\caption{Radial profiles of the surface density for all three tilted simulations. The thicker curves represent the time-averaged surface density, while the thinner curves represent the initial profile at $t=0$. }
\label{fig:sigma}
\end{figure}

Fig.~\ref{fig:sigma} shows radial profiles of the time-averaged surface density (in thicker curves) compared to their initial profiles (in thinner curves) for all three tilted disc simulations. The time averaging is done from $3000\, GM/c^3$ to the end of each simulation. Note that the torus in the initial profile has a density maximum at $9\,GM/c^2$ independent of the thin disc whose surface density is at least an order of magnitude higher than the maximum torus density. Due to this, the maximum density of the torus migrates towards the outer thin disc as the simulation evolves.

\begin{figure*}
\includegraphics[width=\textwidth]{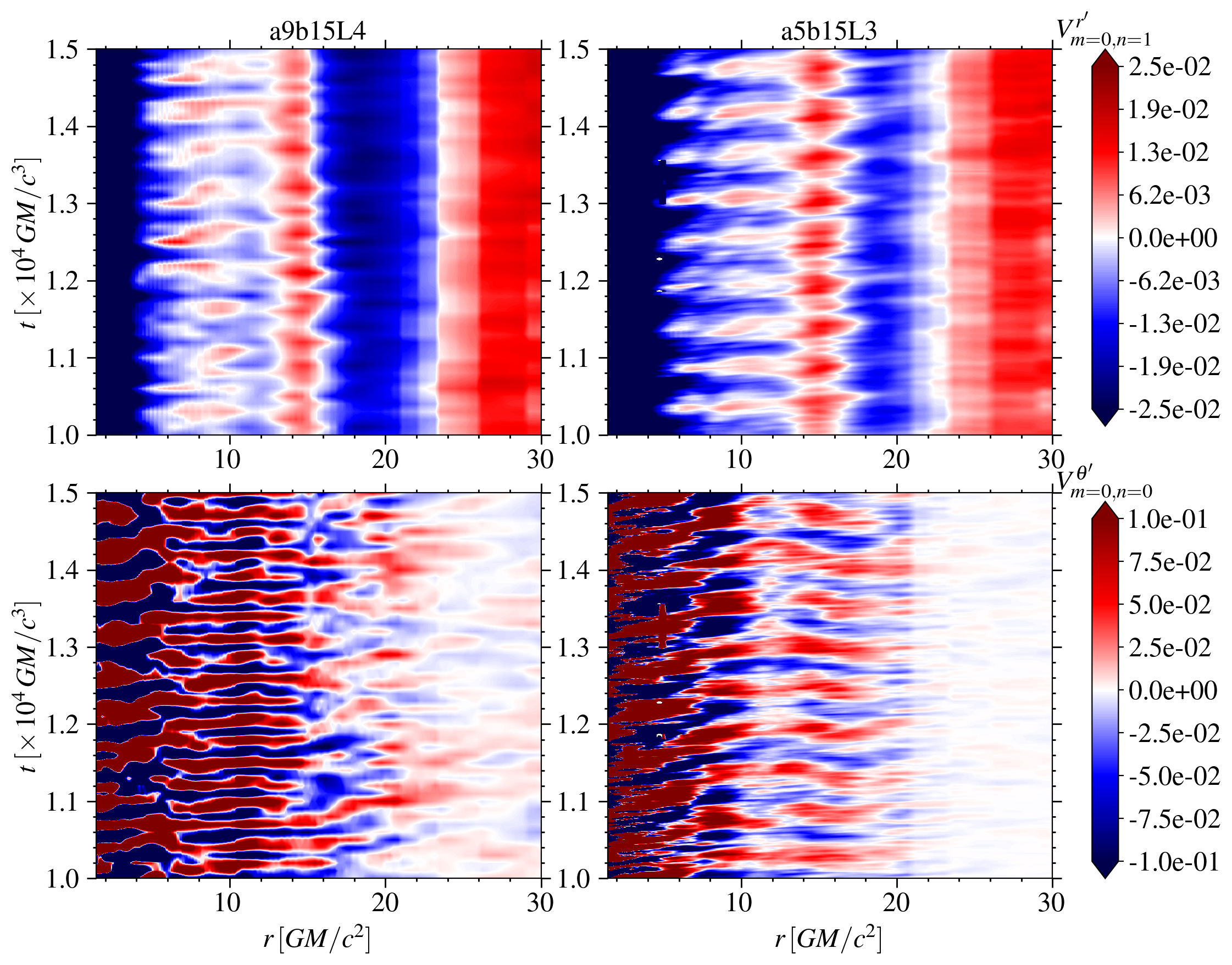}
\caption{Space-time diagrams showing the azimuthally and vertically averaged radial ($V^{r^\prime}$, in top panels) and poloidal ($V^{\theta^\prime}$, in bottom panels) velocities for simulations a9b15L4 (left column) and a5b15L3 (right column). To illustrate the $m=0,n=0$ modes in $V^{\theta^\prime}$, we consider vertical averaging over $\pm 25^{\circ}$ around the mid-plane, while for the $m=0,n=1$ modes in $V^{r^\prime}$, we consider only the upper half of the disc. }
\label{fig:vrvth}
\end{figure*}

To further substantiate the presence of the reported radial and vertical oscillations in Section~\ref{sec:epicyclic}, we present spacetime diagrams of the azimuthally and vertically averaged $V^{r^\prime}$ (top panels) and $V^{\theta^\prime}$ (bottom panels) for simulations a9b15L4 and a5b15L3 in Fig.\ref{fig:vrvth}. To emphasize the $m=0,n=0$ modes observed in Fig.\ref{fig:vth_modepsd}, we employ azimuthal averaging over the entire $2\pi$ and vertical averaging over the disc spanning $\pm 25^{\circ}$ around the midplane. For $V^{r^\prime}$, vertical averaging considers only the upper half of the disc, specifically $\theta \in [65^{\circ}, 90^{\circ}]$. The oscillations around $\sim 0.002,c^3/GM$ are clearly discernible in simulation a5b15L3, while oscillations at relatively higher frequencies ($\sim 0.004,c^3/GM$) are more pronounced in simulation a9b15L4, particularly in $V^{\theta^\prime}$, at least for the time period depicted in this plot. 

\label{lastpage}
\end{document}